%% file: main.tex
\documentclass[a4paper,11pt]{article}
\usepackage{jheppub} 
\newif\ifdraft

\usepackage{hyperref}
\usepackage[dvipsnames]{xcolor}
\usepackage{multirow}

\definecolor{Purple}{rgb}{1.0, 0.0, 1.0}

\definecolor{Pink}{rgb}{1.0, 0.56, 0.68}

\ifdraft
\usepackage{lineno}
\linenumbers
\else

\fi

\title{\boldmath Characterization of the PADME positron beam for the X17 measurement}

\author[a]{S. Bertelli,}
\author[a]{F. Bossi,}
\author[a]{B. Buonomo,} 
\author[a]{R. De Sangro,} 
\author[a]{C. Di Giulio,} 
\author[a,g]{E. Di~Meco,}
\author[b]{K. Dimitrova,}
\author[a]{D. Domenici,} 
\author[c]{F. Ferrarotto,}
\author[a]{G. Finocchiaro,} 
\author[a]{L.G. Foggetta,} 
\author[e]{A. Frankenthal,}
\author[a]{M. Garattini,} 
\author[b,f]{G. Georgiev,}
\author[a]{P. Gianotti,}
\author[b]{S. Ivanov,}
\author[b]{Sv. Ivanov,}
\author[b,a]{V. Kozhuharov,}
\author[c]{E. Leonardi,}
\author[d,c]{E. Long,}
\author[a,g]{M. Mancini,}
\author[d,c]{G.C. Organtini,}
\author[d,c,1]{M. Raggi,\note{Corresponding author.}} 
\author[a]{I. Sarra,}
\author[b]{R. Simeonov,}
\author[a]{T. Spadaro,} 
\author[a]{E. Spiriti,} 
\author[c]{P. Valente,}
\author[c]{A. Variola,}
\author[a]{E. Vilucchi}
\affiliation[a]{INFN Laboratori Nazionali di Frascati, Via E. Fermi 54, 00044 Frascati}
\affiliation[b]{Faculty of Physics, University of Sofia ``St. Kl. Ohridski'', 5 J. Bourchier Blvd., 1164 Sofia}
\affiliation[c]{INFN Roma1, p.le Aldo Moro 5, 00185 Roma}
\affiliation[d]{Physics Department, ``Sapienza'' Universit\`a di Roma, p.le Aldo Moro 5, 00185 Roma}

\affiliation[e]{Department of Physics, Princeton University, P.O. Box 708, Princeton, NJ 08544-0708}
\affiliation[f]{INRNE, Bulgarian Academy of Science}
\affiliation[g]{Department of Physics, ``Tor Vergata'' Universit\`a di Roma, Via della Ricerca Scientifica 1, 00133 Roma}

\emailAdd{mauro.raggi@roma1.infn.it}

\abstract{
This paper presents a detailed characterization of the positron beam delivered by the Beam Test Facility at Laboratori Nazionali of Frascati to the PADME experiment during Run III, which took place from October to December 2022. It showcases the methodology used to measure the main beam parameters such as the position in space, the absolute momentum scale, the beam energy spread, and its intensity through a combination of data analysis and Monte Carlo simulations.
The results achieved include an absolute precision in the momentum of the beam to within $\sim$ 1-2 MeV$/c$, a relative beam energy spread below 0.25\%, and an absolute precision in the intensity of the beam at the level of 2\% percent.
}

\begin{document}
\maketitle
\flushbottom
\section{Introduction}
 
\bigskip
The Positron Annihilation into Dark Matter Experiment (PADME) has been designed with the primary goal to look for a signal of a dark photon~\cite{Holdom:1986eq}. PADME can investigate the existence of different feebly-interacting particles (FIPs) produced in the interaction of a positron beam with a thin diamond target~\cite{PADME:2022fuc}. These particles are predicted by new-physics models beyond the Standard Model (SM) solving the dark-matter problem in scenarios alternative to the WIMP paradigm~\cite{Antel:2023hkf}. 

More and more experiments, all over the world, are involved in the hunt for FIPs, by exploiting different techniques and detectors. In 2016, an experiment at the ATOMKI institute of Debrecen in Hungary studying the de-excitation of $^8$Be via Internal Pair Creation, reported an anomaly that might represents the first evidence of a FIP ~\cite{Krasznahorkay:2015iga}. The opening angle of the $e^+ e^-$ pairs produced was found to be not consistent with the expectation of SM. The disagreement could be explained by an additional de-excitation process, occurring through an intermediate particle state of mass $\approx$ 17 MeV, now named X17. Following this first evidence, the same collaboration confirmed with other measurements on different nuclei the observed signal~\cite{Krasznahorkay:2021joi,Krasznahorkay:2022pxs}. On a parallel line, all the theoretical attempts to justify this anomaly within the Standard Model failed \cite{Zhang:2017zap}\cite{Alves:2023ree}.
PADME is in the unique position of being able to verify the particle hypothesis, by trying to produce resonantly the X17 via the annihilation of its positron beam with the electrons of the target~\cite{Nardi:2018cxi}. The data taking performed by PADME in winter 2022 (Run III) was thus designed with this goal.

The process PADME is investigating is $e^+e^- \to X17 \to e^+e^-$. The X17 production would increase the number of $e^+e^-$ final states detected by the experiment compared to the pure Bhabha scattering background, particularly for center-of-mass energy values near 17 MeV \cite{Darme:2022zfw}.
During Run III PADME collected data at 47 different beam energy values ranging from 262 MeV to 299 MeV, corresponding to a mass interval from $\sim$16.4 to $\sim$17.5 MeV. 
The data analysis focuses on accurately determining the number of cluster pairs in the calorimeter at each energy point and comparing them.

The number of pairs of observed clusters corresponds to the number of final states $e^+e^-\to e^+e^-$ or $e^+e^- \to \gamma \gamma$. If X17 exists, the excess of $e^+e^-$ generated by its decays will significantly alter the measured number. 
The inclusion of the $\gamma\gamma$ final state does not affect much the sensitivity, as its contribution counts only for 20\% of the total. To better compare the number of collected pairs on a run-by-run basis, the number is normalized to $N_{PoT}$, representing the number of Positron-on-Target. The actual observable becoming: $\frac{N_{2cl}}{N_{PoT}}$. In this scenario, the correct and stable measurement of $N_{PoT}$ is of central importance. Additionally, as demonstrated in \cite{Darme:2022zfw}, the production rate of X17 depends on the beam energy spread, which also needs to be measured carefully. 

This paper describes in detail how the positron beam provided to PADME during Run III has been monitored in position and precisely characterized in terms of absolute momentum, energy spread, and intensity. 
The paper is organized as follows: Section~\ref{sec:Beam} describes the BTF beam line and the energy scan technique adopted during Run III; Section~\ref{sec:BeamQual} illustrates the analysis performed by combining experimental data and Monte Carlo simulation and finally reports the values obtained for the absolute beam momentum, the beam energy spread, and the beam intensity; Section~\ref{sec:BeamMon} describes how the position and intensity of the beam spot were monitored during the data taking; Section~\ref{sec:Lumi} illustrates how the luminosity measurement is performed; Section~\ref{sec:Conc} draws some conclusions. 

\input{Experiment}





\bibliographystyle{JHEP}
\bibliography{bibliography.bib}



\end{document}

%% file: Experiment.tex
\section{The LNF Beam Test Facility}
\label{sec:Beam}

\begin{figure}[!h]
\centering
\includegraphics[width=\textwidth]{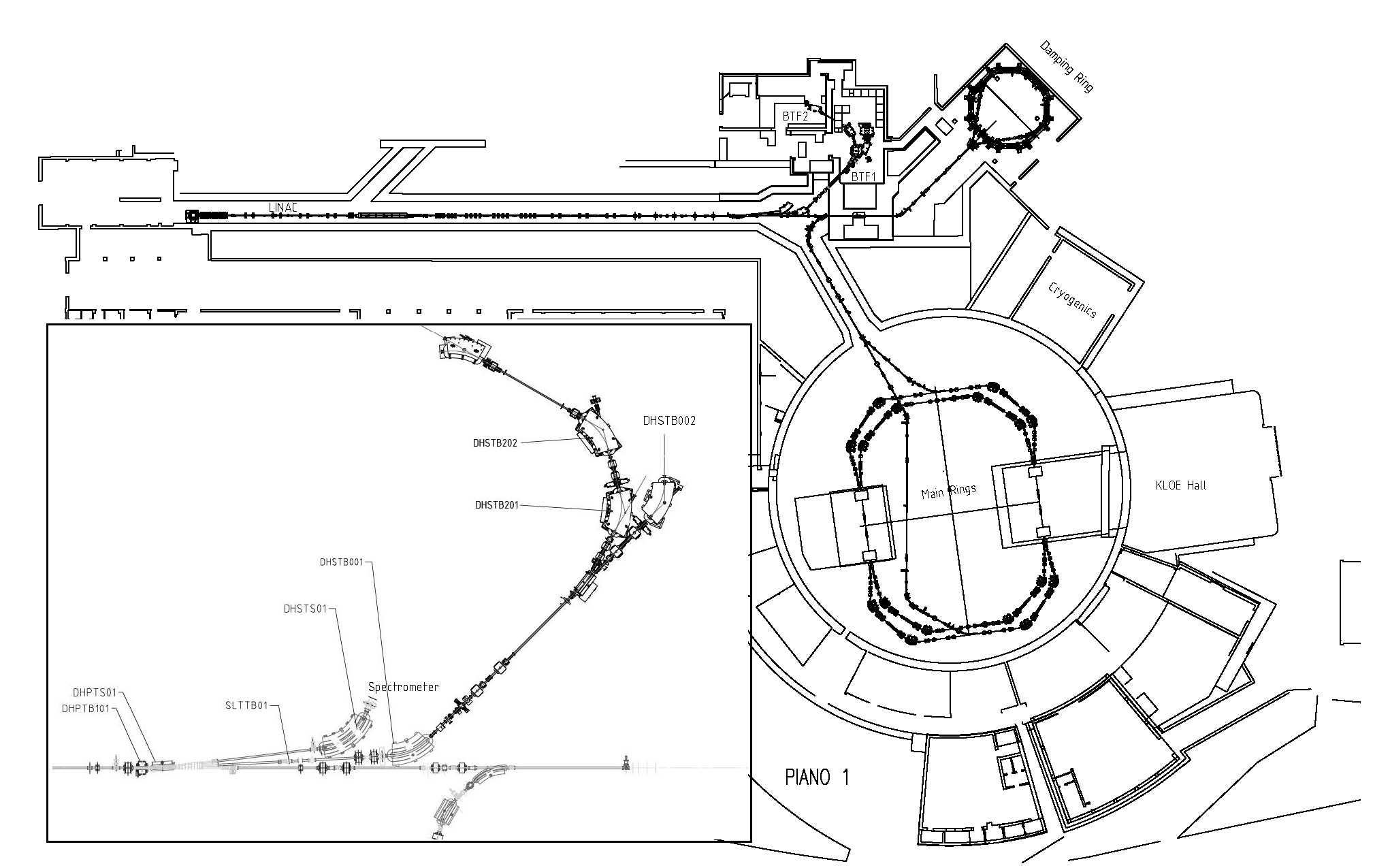}
\caption{The DA$\Phi$NE accelerator complex. In the inset, the detail of the Beam Test Facility transfer line.}
\label{fig:dafnecomplex}
\end{figure}

The LNF Beam Test Facility (BTF) \cite{Ghigo:2003gy} consists of two experimental halls (BTF-1, BTF-2) where beams of electrons/positrons of different energy and intensity are available. 
They are provided by the S-BAND (2856 MHz) LINAC of the DA$\Phi$NE complex~\cite{DAFNE_site} able to produce bunched beams of electrons/positrons with a maximum repetition rate of 50 Hz. In standard operation, the LINAC accelerates particles up to 510 MeV for the DA$\Phi$NE collider: electrons are generated by a 120 KV triode gun, while positrons are produced by the electrons hitting a W-Re target of two radiation lengths (positron converter) located after the first 5 accelerating sections. 
Before being injected in the DA$\Phi$NE ring, electrons/positrons are stacked and damped in a small accumulator in $<13$~ns long bunches.
As an alternative, beam bunches from the LINAC can be extracted to the BTF beam-line. 
At the end of the LINAC a three-way vacuum pipe and a couple of fast ramping ($<10$~ms) pulsed dipoles allow switching the beam pulses away from the straight transfer-line to the accumulator ring. 
The DHPTB101 dipole steers the beam to the 3$^\circ$ line in the BTF channel, while the DHPTS001 deflects the beam to the 6$^\circ$ line towards a high dispersion line and a SEM Hodoscope for the spectroscopic measurement of the LINAC primary beam.
A schematic view of the DA$\Phi$NE accelerator complex, with the detail of the three-way switch-yard in the inset, is shown in Fig.~\ref{fig:dafnecomplex}. 

The maximum energy that electrons/positrons can reach at the LINAC exit is 750/550 MeV with a typical current of 180/85 mA/bunch \cite{Buonomo:2015pba}. 

The Beam Test Facility has been designed to allow the use of LINAC electrons/positrons for detector tests and/or beam-studies. The beam characteristics (spot size, divergence, momentum resolution) depend on the user request in terms of multiplicity (number of particles/bunch) and energy. Furthermore, the energy range, pulse duration,
and duty cycle can be limited during the DA$\Phi$NE collider operation.
\par

The BTF foresees two different operation modes:
\begin{itemize}
    \item[-]{\bf dedicated mode} 
    The LINAC is not employed for injecting electrons and positrons into the DA$\Phi$NE collider. In this mode, there is complete flexibility in the beam parameters, particularly with regard to the beam bunch length.
    
    \item[-]{\bf parasitic mode} The BTF line receives only the LINAC pulses not used for the injection into the DA$\Phi$NE rings. In this configuration there are strong limits on the achievable beam parameters, due to the energy and timing requirements mentioned before.
\end{itemize}
\begin{table}[!h]
\centering
\caption{Main beam parameters in the different BTF operation modes.}
\label{tab:BTFparam}
{\small
\begin{tabular}{p{2cm}p{2cm}p{2cm}p{2cm}p{2cm}} 
\hline
\multirow{2}{*}{\textbf{Parameters} } & \multicolumn{2}{c}{\textbf{Parasitic mode} } & \multicolumn{2}{c}{\textbf{Dedicated mode}}  \\ 
\cline{2-5}
                                      & \textbf{Secondary beam} & \textbf{Primary beam}                                                 
                                      & \textbf{Secondary beam} & \textbf{Primary beam}                                             \\ 
\hline
Electrons or Positrons                & Selectable at BTF    & Depending on DA\ensuremath{\Phi}NE  & Selectable at BTF    & Selectable at LINAC                                                 \\
\hline
Energy          & 30-500  MeV         & 510   MeV       & 30-700               & \begin{tabular}[c]{@{}c@{}}167-750 (e$^-$)\\250-550 (e$^+$)\end{tabular}  \\
\hline
Energy spread       & 0.5\%\hspace{4em}at 500 MeV       & 0.5 \%  & 0.5\%\hspace{4em}at 500 MeV       & 0.5\%  \\
\hline
Intensity                             & 1-10$^5$             & 10$^7$-1.5$\times 10^{10}$                                                    & 1-10$^5$             & 1-3$\times 10^{10}$                                              \\
\hline
Pulse width                      & \multicolumn{2}{c}{10  ns}  & \multicolumn{2}{c}{1.5-300 ns}    \\
\hline
Rate                  & \multicolumn{2}{c}{\begin{tabular}[c]{@{}c@{}}1-49 Hz\\(depending from DA\ensuremath{\Phi}NE injection)\end{tabular}} & \multicolumn{2}{c}{1-49 Hz}                                                                   \\
\hline
\end{tabular}
}
\end{table}
The PADME physics program requires to operate the LINAC in the {\it dedicated mode} and with a proper setup of the BTF beam line. 

The positron beams into the BTF can be obtained through two distinct approaches: either through production at the LINAC positron converter, referred to as the {\it primary positron beam}, or via generation on a secondary target installed on the BTF line, known as the {\it secondary positron beam}. For PADME data collection, the {\it primary positron beam} is preferred due to its capability to minimize background in the apparatus~\cite{PADME:2022ysa}. A fine tuning of the BTF beam energy is obtained acting on the current of a 45$^\circ$ static dipole (DHSTB001), more details on the procedure are given in next subsection.

Table \ref{tab:BTFparam} summarizes the beam parameters that can be obtained in the different operation modes. 
\par
More details about the LNF BTF can be found in~\cite{Foggetta:2021gdg}.

\subsection{Beam parameters and energy selection scheme}
\begin{figure}[h]
\centering
    \includegraphics[width=\textwidth]{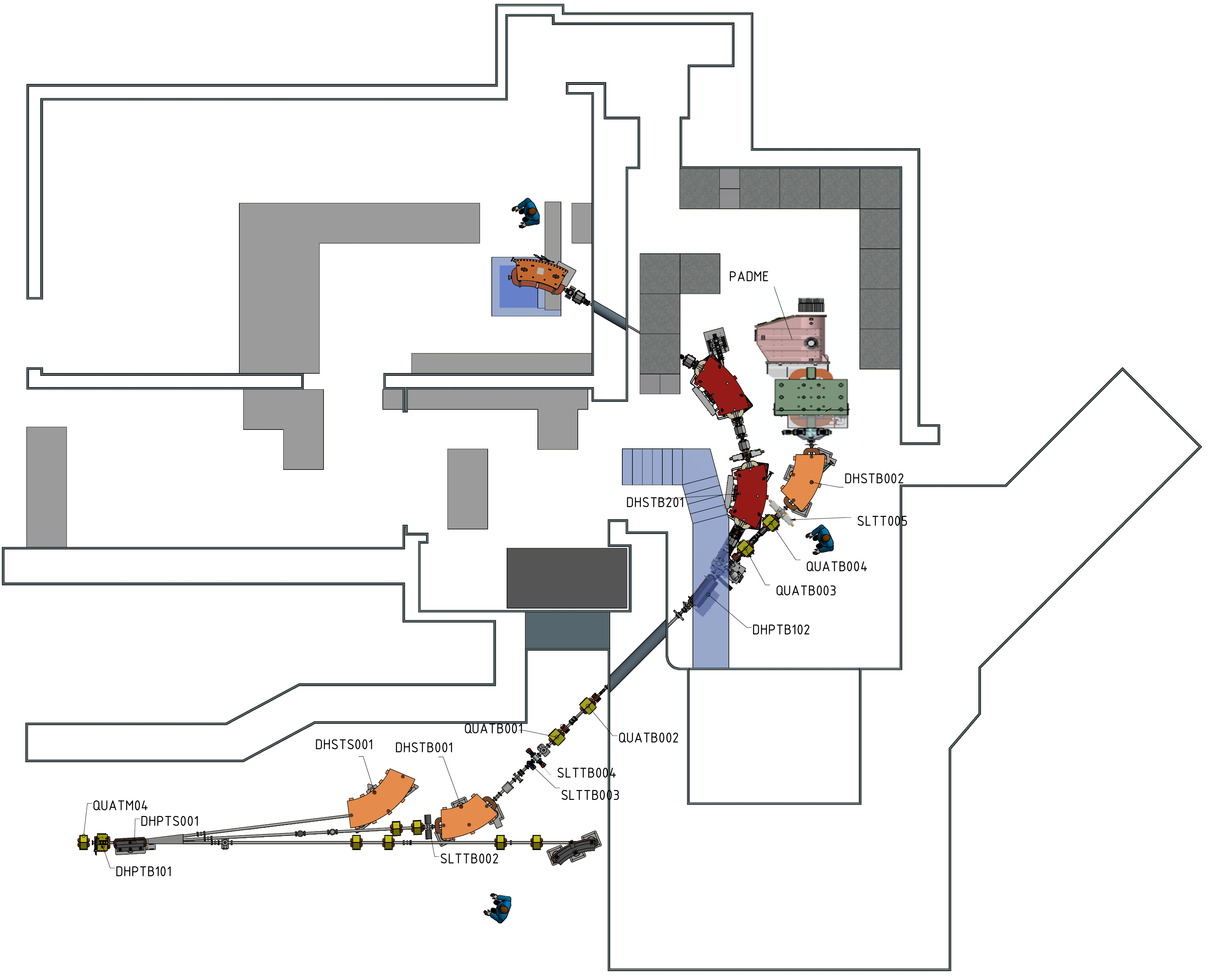}
    \caption{PADME beam line. All the elements of the momentum selection scheme are indicated (see text for more details).}
    \label{fig:selector}
\end{figure}

The positron beam provided to the PADME experiment can be precisely tuned in terms of energy, dimension, and intensity.
The momentum selection scheme of the BTF is illustrated in Fig.~\ref{fig:selector}. It involves properly adjusting the current of dipole DHSTB001. This is a 45$^\circ$ sector magnet slightly displaced and rotated in order to compensate the 3$^\circ$ bending introduced by the pulsed extracting magnet DHPTB101. It operates at a current suited to produce (for a given energy setting $E_{sel}$)  the additional 42$^\circ$ needed to drive the beam in the beam pipe exiting the LINAC tunnel wall at 45$^\circ$, and entering the BTF experimental areas.

The energy spread around $E_{sel}$ is contingent upon the aperture of the two collimator systems in the horizontal plane: the SLTTB004 collimator, situated downstream of the selecting dipole, absorbs particles diverging significantly from the central trajectory owing to the dipole's chromatic effect. Conversely, the SLTTB002 collimator, positioned immediately upstream of the dipole, restricts the spread of the entrance angle within the magnetic field and confines the horizontal position.

Considering the typical optics used during PADME Run III for the collimator apertures, a relative energy spread below 0.5\% is expected.

Downstream of DHSTB001, the beam optics is controlled by two focusing-defocusing quadrupole doublets: QUATB001-002, situated in the LINAC area, and QUATB003-004 positioned inside the BTF-1 experimental hall. In addition to the pair of collimators SLTTB002 and SLTTB004 operating on the horizontal axis, a pair of collimators labelled SLTTB001 and SLTTB003 operates on the vertical axis.

The beam intensity, specifically the number of Positrons On Target per bunch ($N_\mathrm{POT}$), can be adjusted by fine-tuning the optics of the LINAC and/or of the transport line. A further beam intensity control is achieved by using the aforementioned pairs of collimators as beam scrapers.

During Run III, the average beam intensity was $\sim (2.5-3.5)\times10^3$ positrons/bunch. Along the whole data taking, the PADME lead glass detector provided a measurement of $N_\mathrm{POT}$ for each bunch. To assess its associated systematic uncertainty, independent detectors were used in specific configurations:
\begin{itemize}
\item a FitPix solid state detector, part of the BTF instrumentation, provided a flux measurement after any new beam-energy point setting;
\item a lead glass detector, also part of the BTF instrumentation, was used to cross-check the measurement provided by the FitPix in specific controlled conditions.
\end{itemize}

The use of this information and the methodology for the precise measurement of the PADME luminosity are detailed in Section~\ref{sec:BeamQual}.

\section{The PADME RUN III positron beam}

\label{sec:BeamQual}

The PADME X17 dataset, denoted as Run III, was acquired between October and December 2022. It was collected at various positron beam energy values, overall 47 distinct center-of-mass energy points. This scanning process covers the center-of-mass region identified by the Atomki collaboration as significant for observing the hypothetical X17 particle.
The energy scan spanned from 261.7~MeV to 299.3~MeV, corresponding to a mass range from 16.35 MeV to 17.48~MeV. 
In addition to the primary data-set, two supplementary data samples were also collected. The first one was acquired at a higher beam energy of 402 MeV, while the other spanned five energy points below the resonance, ranging from 205 MeV to 211 MeV.

Using the BTF magnetic transport line, we selected various positron energies by adjusting only the current of the DHSTB001 magnet (see Fig.~\ref{fig:selector}). This technique allowed a reduction of the number of changes for the LINAC from $\sim$~50, one per each energy value, to around 15, thus ensuring more stable and uniform beam conditions. 

\subsection{Absolute beam momentum setup}

The absolute LINAC's beam momentum is given by the LINAC's beam diagnostic instruments which measure its deflection in the horizontal plane at the exit of the dipole positioned at the LINAC's end within the spectrometer line as depicted in Fig. \ref{fig:selector}. 
This is done diverting one bunch per second from the splitting magnet (DHPTB101) into the afore mentioned spectrometer line. 
Additionally, an independent measurement is performed thought the magnetic field of the DHSTB001 dipole monitored using a Hall probe, along with its current provided by the magnet power supply.

The PADME experiment records the value of the DHSTB001 magnet current, magnetic field, and the beam energy value provided by the LINAC's beam diagnostics on a run-by-run basis. 
In addition, the measured values of the magnetic fields of all dipoles, the quadrupole gradients, and the collimator's apertures are recorded and used as input parameters for the PADME Monte Carlo (MC) simulation. 
The nominal energy accepted by the BTF transport line, $E_{Sel}$ is calculated from the measured current of the DHSTB001 magnet, using its nominal magnetic length and the experimental field/current excitation curve of the dipole \cite{Ghigo:2003gy}. 
The values of the measured magnetic field from an Hall probe at the dipole exit have been compared with those obtained from the measured magnet current $I$ and the excitation curve in \cite{Ghigo:2003gy}:
\begin{equation}
    B\,[G]=28.42\times I\,[A]+16.22
    \label{eqn:exCurve}
\end{equation}
The observed difference is within 20--30~G over the entire scan period. This can be due to a different hysteresis curve of the magnet during PADME Run III with respect to the conditions in~\cite{Ghigo:2003gy}, thus affecting the constant term in Eq.~\ref{eqn:exCurve}. The magnetic field applied during Run III is in the range of 4000-5000~G. The 30 G difference observed corresponds to a relative error on the absolute momentum scale below 1\%.
To further assess the systematic error in the absolute energy scale, we compared the value of the selected energy provided by the BTF diagnostics to the one obtained from the beam line MC simulation described in~\cite{PADME:2022ysa}.

Using as MC inputs the magnetic field values and the collimator's positions for a number of different beam configurations, the energy resulting from the beam line MC agrees with the one provided by the BTF instrumentation to the level of $\sim$~1 MeV. Changing the magnetic field value on the transport line by 30~G we induce a difference of $\sim$1.5~MeV on the beam energy, which translates into a 40~KeV shift in the center of mass energy.

\subsection{Beam energy spread}

The  dipole magnet DHSTB001 introduces a chromatic dispersion, creating a correlation between the positron beam energy and its angular deflection. 
The relative energy spread $\Delta E/E$ is therefore directly proportional to the relative angular dispersion $\Delta \phi/\phi$.
The drift length following the bending magnet transforms the energy-angle into a energy-$x$ correlation, where $x$ represents the horizontal beam deflection. The downstream collimator slit SLTTB004, with an aperture of h positioned at distance L1=1.4750~m, allows only particles within a specific momentum range to pass through.\cite{Ghigo:2003gy}:
\begin{equation}
    \frac{\Delta E}{E} = \frac{h}{2\rho} + \sqrt{2} \left(\frac{R_x}{L_1} + \frac{H}{2L_1}\right)
    \label{eqn:SannValente}
\end{equation}
In Tab. \ref{tab:Coll_Pos} the actual positions of the BTF beam line collimators during Run III are listed.
\begin{table}[h]
    \centering
    \begin{tabular}{|l|c|}
\hline
Collimator name & Aperture\\
\hline
SLTB2 (H) &  1.5 mm\\
SLTB3 (V) & 4.3 mm\\
SLTB4 (h) &  2.1 mm\\
 \hline        
    \end{tabular}
    \caption{Standard collimator's position during PADME Run III.}
    \label{tab:Coll_Pos}
\end{table}

The value of the collimators gaps are H=1.5 mm and h=2.1 mm.  
For the BTF line $\rho=1.723$~m. Assuming $\sigma_x$ in the range 2--4~mm, one obtains $\Delta E/E=$0.22--0.36\%.

\begin{figure}[!h]
\centering
\includegraphics[width=0.7\linewidth]{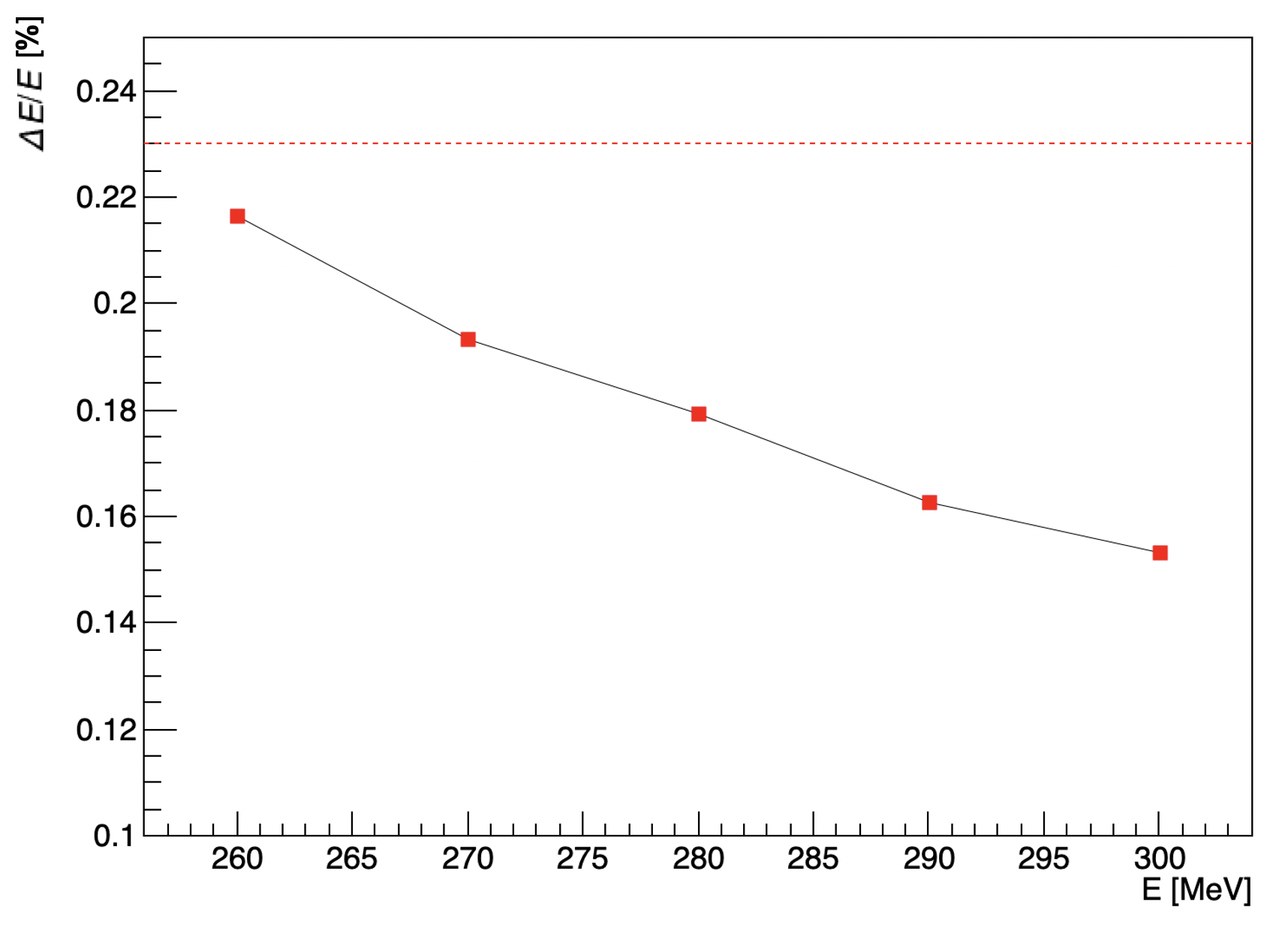}
\caption{Simulated values of $\frac{\Delta E}{E}$ as function of the beam energy in MeV.}
\label{fig:MCESpread}
\end{figure}

A substantial correction to the beam energy spread, not accounted for in \ref{eqn:SannValente}, arises from the beam optics, particularly the FODO quadrupole doublet QUATB001-QUATB002. 
Throughout the entire scan, the beam optics remained almost unchanged. 
Due to the substantial range of beam energies used, we anticipate that the variation of the beam spot size at the downstream collimator plane, attributed to the quadrupole's focusing and defocusing effects, will influence the beam energy spread.
The effect described is not accounted for in the analytical expression (Eq. \ref{eqn:SannValente}), therefore we used beam line Monte Carlo simulation to gain a more accurate insight.
Fig.~\ref{fig:MCESpread} shows the beam energy spread obtained by the MC simulation at the PADME target plane, using the complete Run III beam line setup.
The MC simulation confirms the value of approximately 0.2\%, in line with the analytical estimate, and indicates a mild dependence on the beam energy, as expected. 
Based on the two estimates, we can confidently state that the beam energy spread during PADME Run III was below 0.25\%. The variation observed in the Monte Carlo simulations does not affect much the X17 production rate, as the dominant contribution comes from the electrons motion in the diamond target\cite{Arias-Aragon:2024qji}.

\section{Beam monitoring}
\label{sec:BeamMon}

Although the search for an excess in the cross-section for the process $e^+e^-\to e^+e^-$
can be assumed to be a counting experiment, 
the monitoring of the beam parameters, including the beam spot position and the 
beam multiplicity,
was found to be critical to achieve the physics goal of the experiment.
The acceptance for the $e^+e^-$ events, the reconstruction efficiency, 
and the luminosity correction factors depend on the beam position and its spread. 
PADME employs several detectors to 
trace any change in the beam parameters.

\subsection{PADME Timepix array}
\begin{figure}[h]
  \centering
  \includegraphics[width=0.6\textwidth]{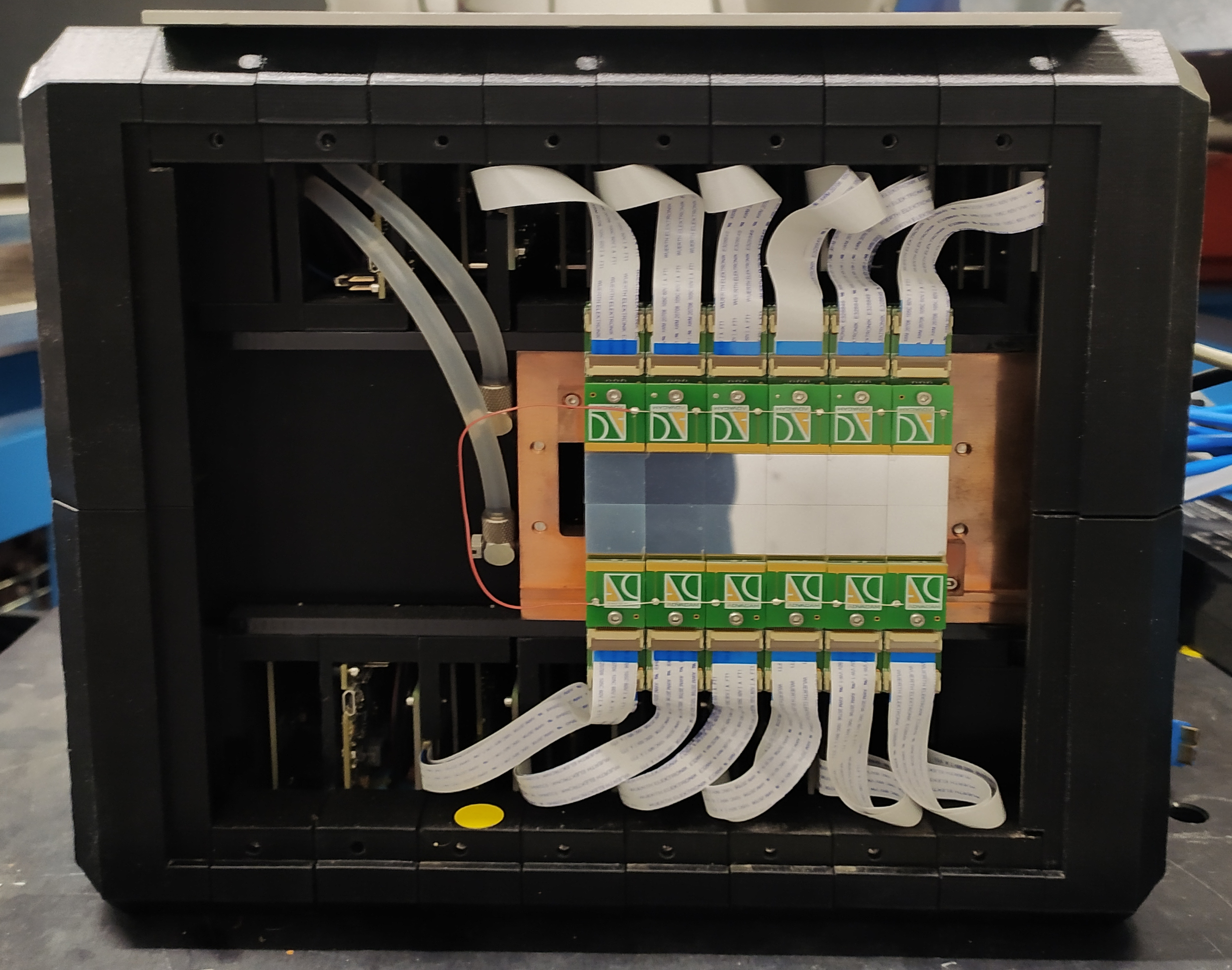}
  \caption{\label{fig:timepix-picture} PADME's Timepix detector without the front cover.}
\end{figure}

A high granularity silicon pixel detector was employed to monitor the
beam parameters at the exit of the PADME experimental setup. The detector consists of an array of 2 $\times$ 6 Timepix3 sensors with a total active sensing area of 28 $\times$ 84 mm (See Fig. \ref{fig:timepix-picture}). PADME's Timepix detector relies on an external 40 MHz clock for time synchronization of the individual chips and is not connected to the PADME trigger distribution system.

A single Timepix3 chip is built from 256 $\times$ 256 pixels and provides measurements of ToA (Time-of-Arrival) and ToT (Time-over-Threshold) for each individual pixel. Each pixel is formed of a 300 \textmu m thick layer of Si and has a square shape with a pixel size px = 55 \textmu m. It can operate in two different modes: frame mode, providing ToA/ToT measurements for each pixel in a predefined interval of time; and data-streaming (DS) mode, providing continuous measurements of ToA/ToT for each pixel with a time resolution of 1.56 ns.

Measurements with PADME's Timepix provide information about the variation of the beam parameters on different scales: on a run level with a run average estimation; measurement of the variation during an individual run; and sensitive to changes of the beam characteristics within a single bunch of positrons ($\sim$ 230 ns), exploiting the data-streaming operating mode.

During Run III, Timepix was operated primarily in the frame mode, 
collecting frames every 2 minutes with acquisition time of 1 s each. 
During the operation, some of the pixels reported missing or corrupted data. 
These pixels were located near the center of the beam spot, 
which introduced a challenge for accurate measurements of the beam characteristics. 
Several techniques to overcome this limitation were developed 
and a Gaussian fit over the upper row of chips was chosen 
to be the most accurate for position sensitivity measurements \cite{timepix:padme-beam}.

\subsection{Beam position}

The Timepix detector is capable to  track the beam position
at the exit of the PADME experiment
across the 
different beam energy values examined.
However, the Timepix detector was not operational during the
full data taking. 

An alternative approach involves reconstructing the center of gravity of the 2-cluster events after the selection process. This method enables the measurement of both the X and Y extrapolated average impact positions of the beam on the front face of the luminometer, with a few mm uncertainty.

\begin{figure}[h]
    \centering
    \begin{minipage}{17.5pc}
      \centering
      \includegraphics[scale=0.3]{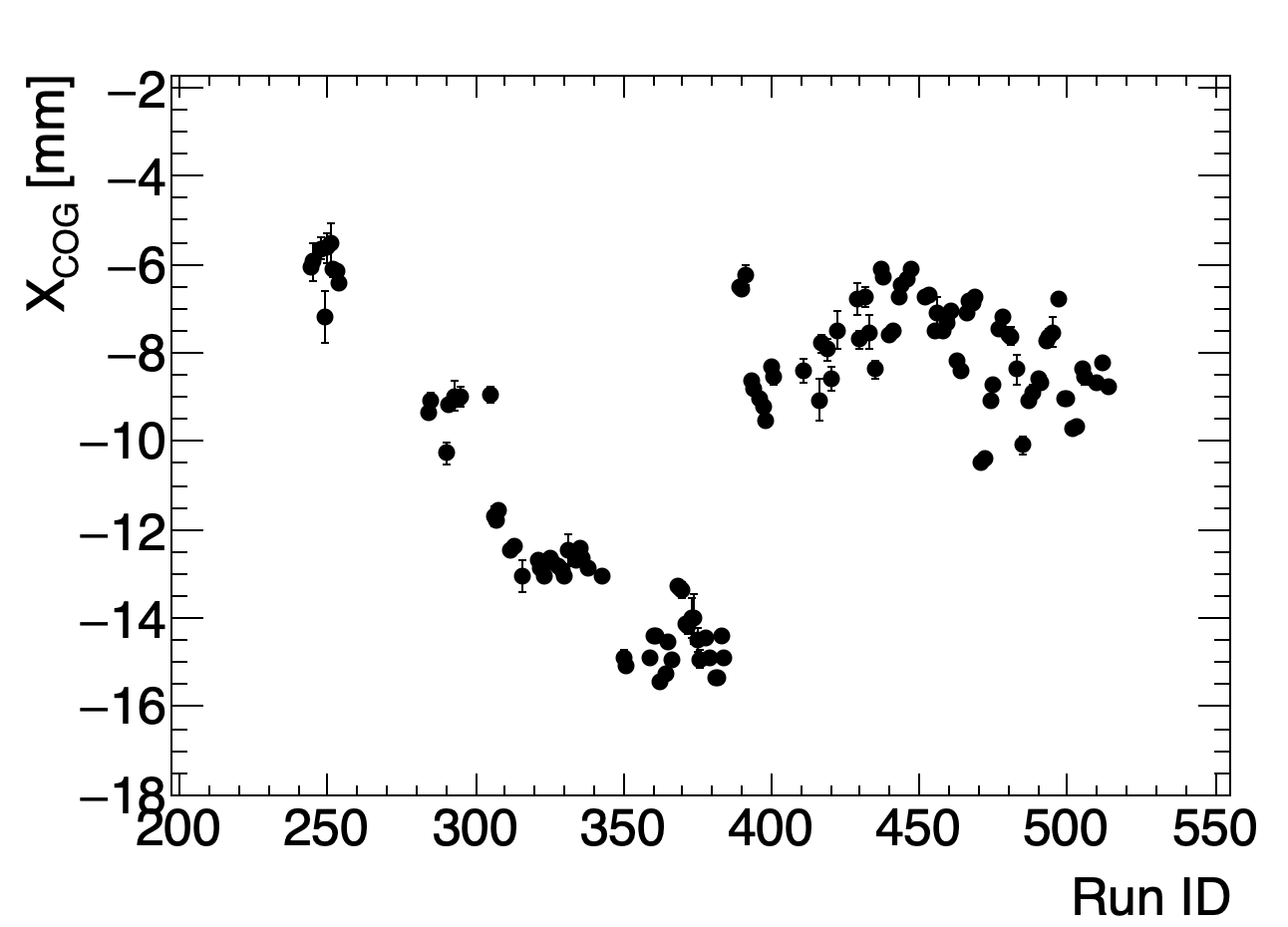}
    \end{minipage} \hfill
    \begin{minipage}{17.5pc}
      \centerline{
        \includegraphics[scale=0.3]{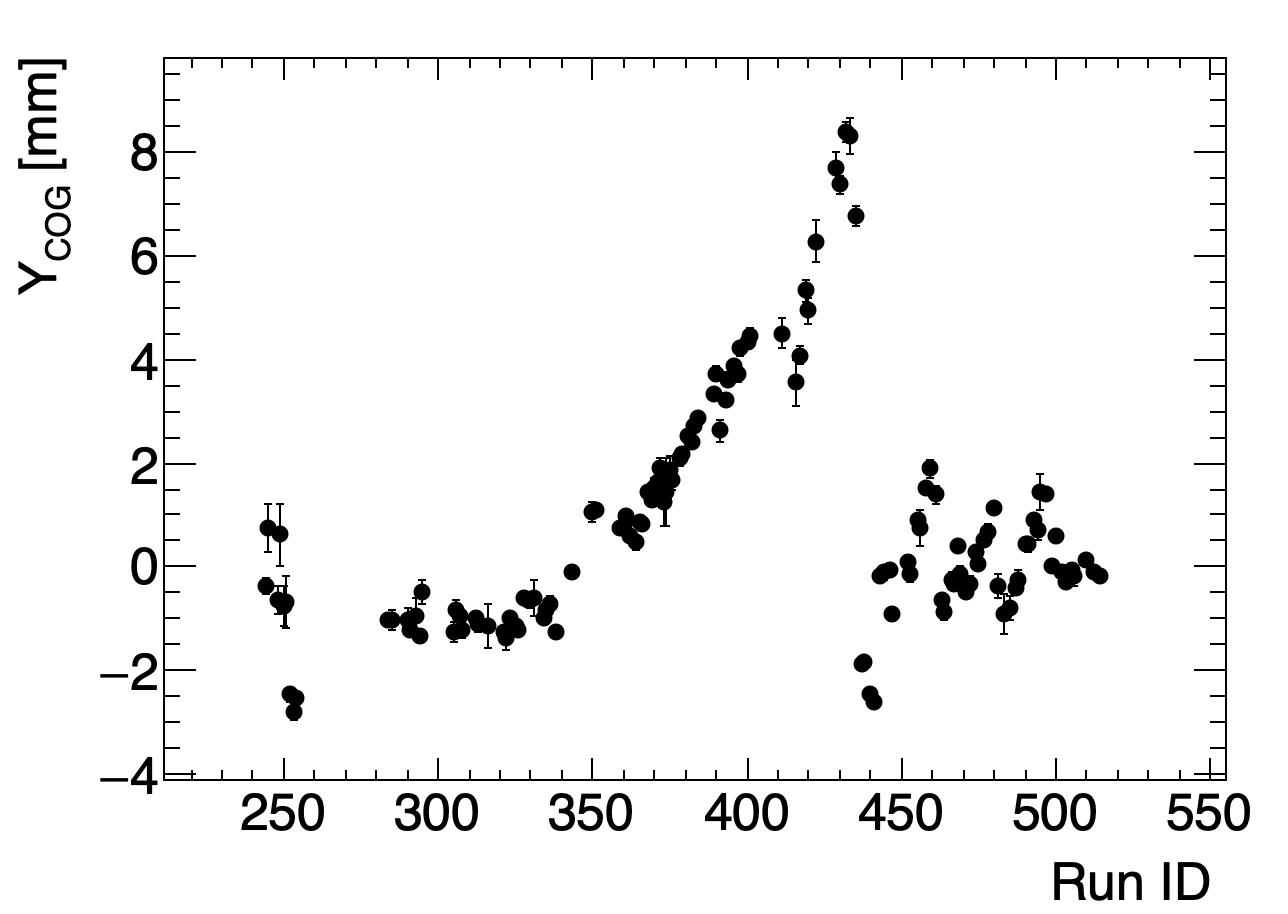}}
    \end{minipage}
    \caption{The beam position at the ECal plane estimated by calculating the average center of gravity of two cluster events 
    for each run.}
    \label{fig:COG_ECal}
\end{figure}

The beam position at the calorimeter plane, as shown in Fig. \ref{fig:COG_ECal}, exhibits fluctuations of the order of 10 mm throughout the data taking period, predominantly in the X coordinate.
The average impact point deviates from the geometrical center by approximately 10~mm. This will induce a systematic correction of the order of percent to be applied to the charge collected by the PADME lead-glass based luminometer, which is placed downstream the calorimeter central hole.
 
\begin{figure}[h]
    \begin{minipage}{17.5pc}
      \centering
      \includegraphics[scale=0.3]{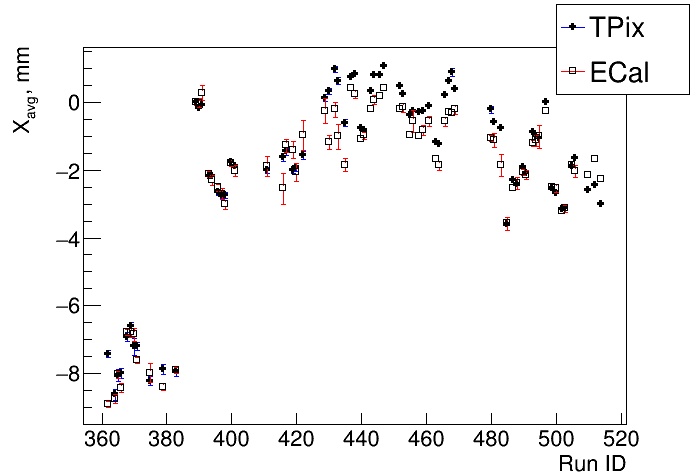}
    \end{minipage} \hfill
    \begin{minipage}{17.5pc}
      \centerline{
        \includegraphics[scale=0.3]{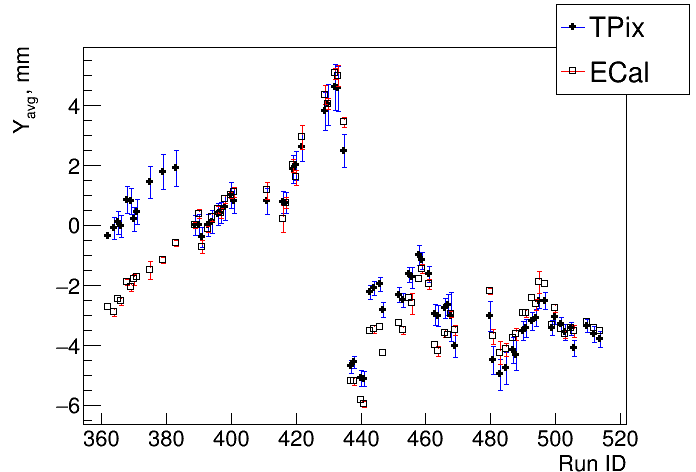}}
    \end{minipage}
     \begin{minipage}{17.5pc}
      \centering
      \includegraphics[scale=0.3]{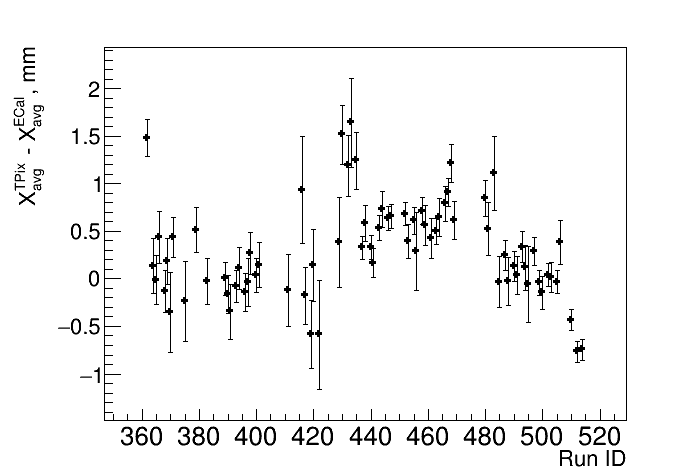}
    \end{minipage} \hfill
    \begin{minipage}{17.5pc}
      \centerline{
        \includegraphics[scale=0.3]{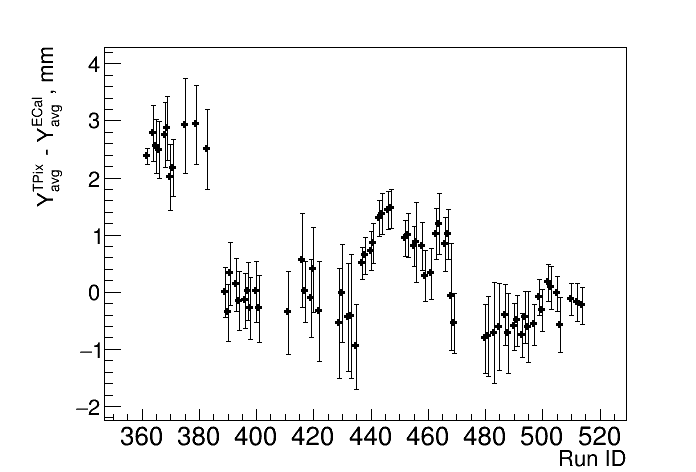}}
    \end{minipage}
    \caption{Comparison between the measurement of the beam center position in X direction (left) and Y direction (right) with the ECal(using the CoG method for two cluster events) and the Timepix 
    (using Gaussian fit on the upper row chips). 
    The bottom plots present the difference between the 
    values obtained with the two methods, 
    showing consistency within 1 mm. 
    The period up to Run ID 384 is with different 
    placement of Timepix3 detector. 
    }
    \label{fig:timepix-vs-ecal-center}
\end{figure}

The measurements of the 
center position of the positron beam with Timepix are compared 
to the Center-of-Gravity measurement of the 2-cluster events in the calorimeter.
The resultant variation of the beam position 
measured with Timepix and with the CoG at the calorimeter is
shown in Fig. \ref{fig:timepix-vs-ecal-center}. 
The presented data spans the period from run 50362 to run 50506
during which the Timepix was operated reliably in frame mode.
As it can be seen, 
measurements agree to the level of better than 1 mm throughout the data taking period considered.
The fluctuations of the beam central position 
at the Timepix detector are 
of the order of O(10) mm 
in both directions.
The 
changes between two consecutive runs are larger in X direction, 
while in Y the shifts are smaller. 
As already stated 
for the beam position measurement with the calorimeter
this is due to the fact that the beam line dipole 
magnets bend the positron bunch mainly in the X direction.

\begin{figure}[h]
    \centering
    \begin{minipage}{17.5pc}
      \centering
      \includegraphics[scale=0.3]{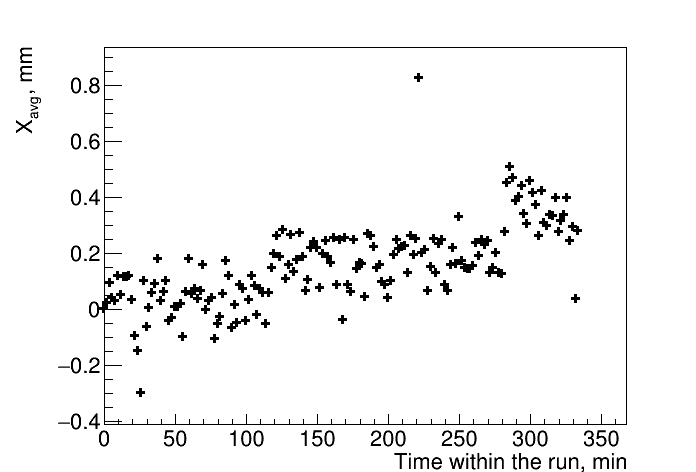}
    \end{minipage} \hfill
    \begin{minipage}{17.5pc}
      \centerline{
        \includegraphics[scale=0.3]{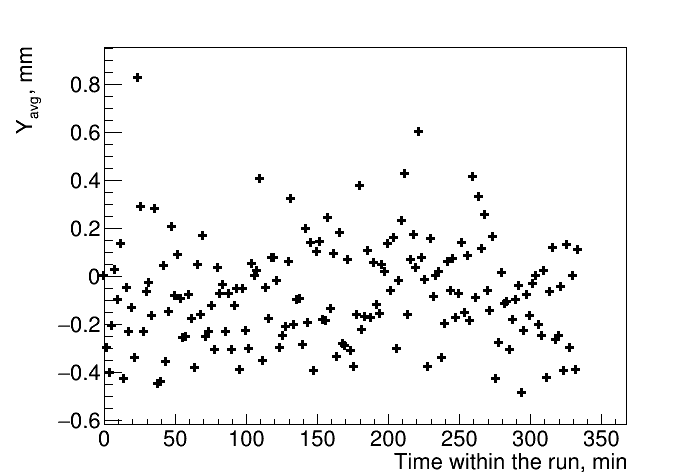}}
    \end{minipage}
    
\ifdraft
    \begin{minipage}{17.5pc}
      \centering
      \includegraphics[scale=0.3]{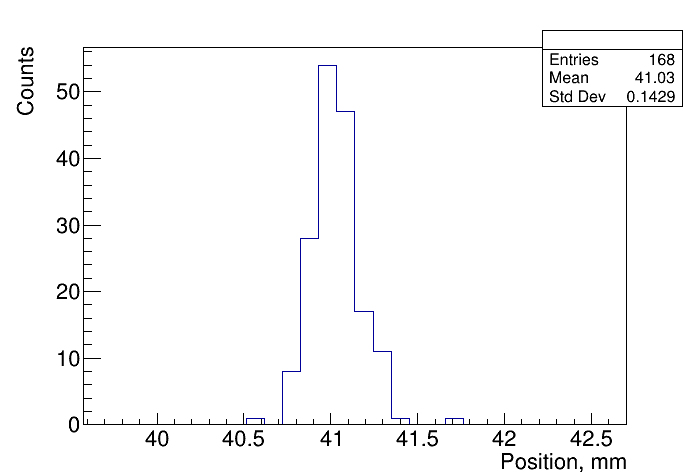}
    \end{minipage} \hfill
    \begin{minipage}{17.5pc}
      \centerline{
        \includegraphics[scale=0.3]{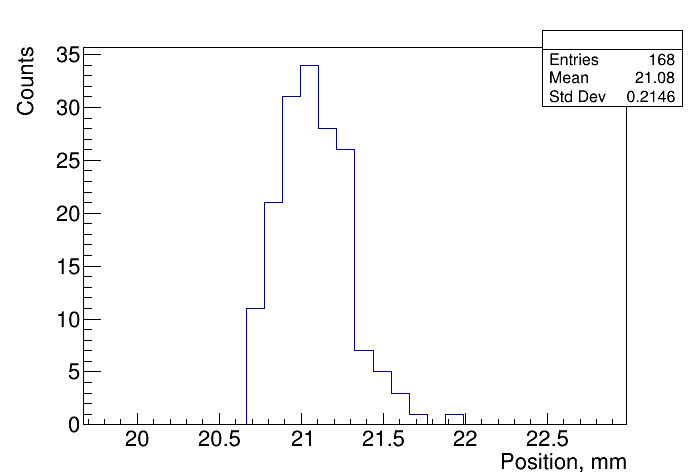}}
    \end{minipage}    
\fi

    \caption{Relative variation   of the beam center position in X direction (left) and Y direction (right) parameters for run number 444 
    with respect to the first recorded Timepix frame. 
    The shift in X direction is of the order of 300~\textmu m. }
    \label{fig:timepix-single-run}
\end{figure}

Due to the frequent recording of the Timepix frames
the collected data allows to follow the variations of 
the beam within a single run with given energy. 
Such monitoring is not possible with the calorimetric 
reconstruction of the center of gravity due to the 
lack of statistics in the two-cluster events sample. 
A measurements of the variation of the beam center position 
both in X and Y
during  run number 444 with energy of the positrons
E~=~289.52~MeV  is shown
in Fig. \ref{fig:timepix-single-run}. 
As can be seen,
even during a single run with a constant beam energy, 
any small shift in the position of the beam can be detected. 
For the presented run, 
the center of the beam was shifting at the end of the run in X direction, 
while in Y it remained stable.
The observed relatively large fluctuations in Y direction
can be attributed to the 
smaller statistics, due to the missing data from chips
in the bottom row of the detector plan and their reflection in the Gaussian fit. 

\begin{figure}[h]
  \begin{minipage}{17.5pc}
    \centering
    \includegraphics[scale=0.3]{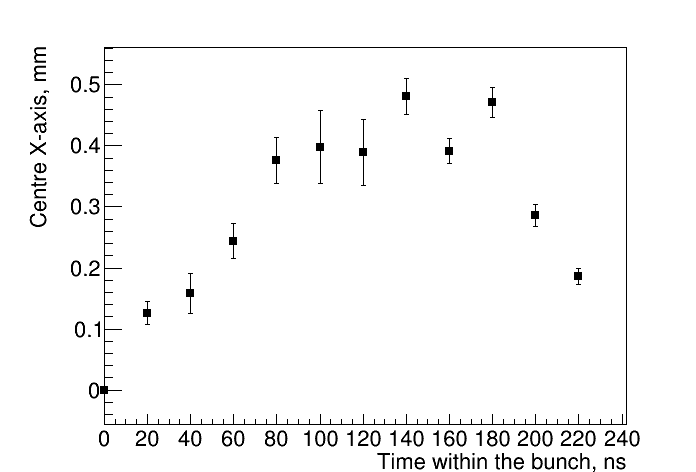}
  \end{minipage} \hfill
  \begin{minipage}{17.5pc}
    \centerline{
      \includegraphics[scale=0.3]{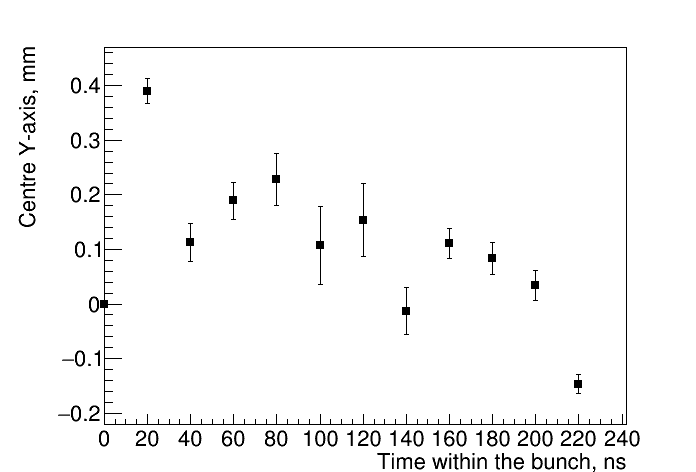}}
  \end{minipage}
  
  \caption{Measurement of the beam center in X direction (left) and Y direction (right) within 
  the bunch with duration of 230 ns, 
  obtained by averaging the information recorded in stream mode for 100 bunches. 
  A small drift of the center position of the positron spot in X was observed, while in Y the beam center remained stable.}
  \label{fig:timepix-single-bunch}
\end{figure}

The total displacement of the observed beam center position during the run is approximately 300 $\mu$m, occurring at a distance of more than 3 m from the PADME target. This corresponds to an angular variation of 
around 100 $\mu$rad, corresponding to a relative 
change in the beam momentum of $\delta p/p \simeq 300 \rm{keV}$. 
This value is negligible compared to the 1-2 mrad beam divergence. 
The data indicates that there is no significant movement of the beam within a run, 
consequently ensuring the stability of the acceptance over time in a single run.
The obtained spread  
of the beam center position 
for run 50444 
is of the order of 150 \textmu m  in X direction 
and 220 \textmu m  in Y direction, 
and is typical for each run during the whole data taking period.

The framework collecting data in the more sophisticated data-streaming (DS) mode 
was completed at the end of Run III and PADME registered data in this mode for tree different beam energies. 
These measurements were used for test and validation of the data acquisition framework and analysis software. 
The DS measurements, unlike the frames, are able to register two or more hits occurring in a single pixel. Moreover, exploiting the time resolution of 1.56 ns, a comprehensive estimation of the beam flux, spread and profile within a single bunch is available.

As shown in Fig. \ref{fig:timepix-single-bunch} obtained using the data recorded 
for a single bunch in data-streaming mode, 
a small drift of the center position of the beam in X was observed, 
while in Y it remained stable during the bunch length of $\sim$ 230 ns. 
The change of the center position, 
taken as the difference between the minimal and maximal value in X direction
is $\sim$ 350 \textmu m. 
This drift was found to be negligible, given the positron beam momentum spread,
and 
confirm
that there is no significant movement of the beam not only within a run, 
but even on a bunch level.

\subsection{Beam spread}

The size of the beam spot is important for the 
control of the beam focusing and for the estimation the energy loss 
in the beam flux measurement. 

\begin{figure}[h]
    \centering
    \begin{minipage}{17.5pc}
      \centering
      \includegraphics[scale=0.3]{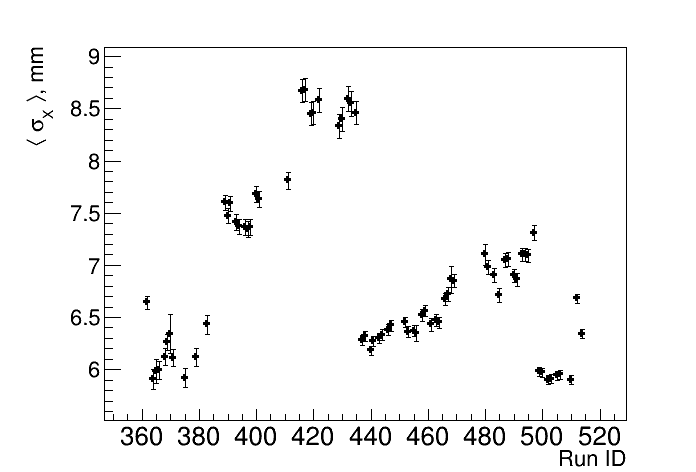}
    \end{minipage} \hfill
    \begin{minipage}{17.5pc}
      \centerline{
        \includegraphics[scale=0.3]{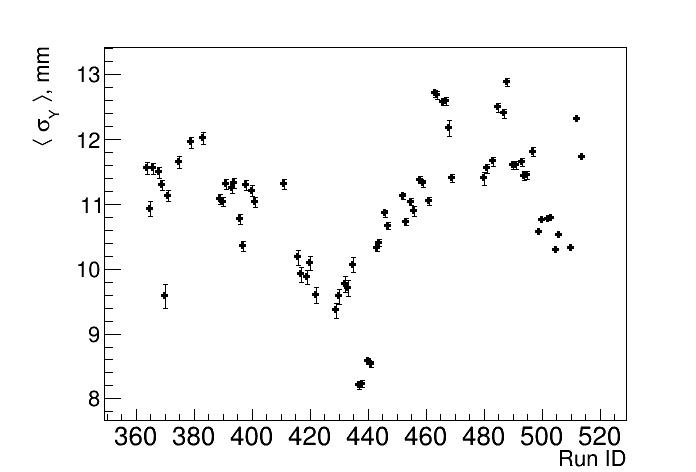}}
    \end{minipage}
    \caption{Variation of the average value of the central position and the spread of the beam spot over the runs. Left is X direction, right is Y. 
    }
    \label{fig:timepix-gauss-all-sigma}
\end{figure}

Using the Timepix, an estimation of the beam size is 
given by the standard deviation of the gaussian 
fits of the spot in X and Y direction. 
The average beam size in X and Y for each run 
is shown in Fig. \ref{fig:timepix-gauss-all-sigma}. 
As can be seen from the distributions, 
the beam size at the exit of the PADME experiment 
exhibits a greater value 
for the sigma in Y direction, $\bar{\sigma_Y} \sim$~11~mm, 
compared to $\bar{\sigma_X} \sim$ 7.5 mm in X direction. 
Using the position of the 
last dipole magnet and assuming that the beam focal 
point is inside the magnet,
the beam size can be translated into a limit of the beam 
divergence, which amounts to $\delta(\varphi)/\varphi \sim 0.002$.
Due to the 
constant currents in the focusing magnets along the positron beam line,
the change of the size of
the beam spot is correlated with the change of the energy of the incident positrons. 

\begin{figure}[h]

    \begin{minipage}{17.5pc}
      \centering
      \includegraphics[scale=0.3]{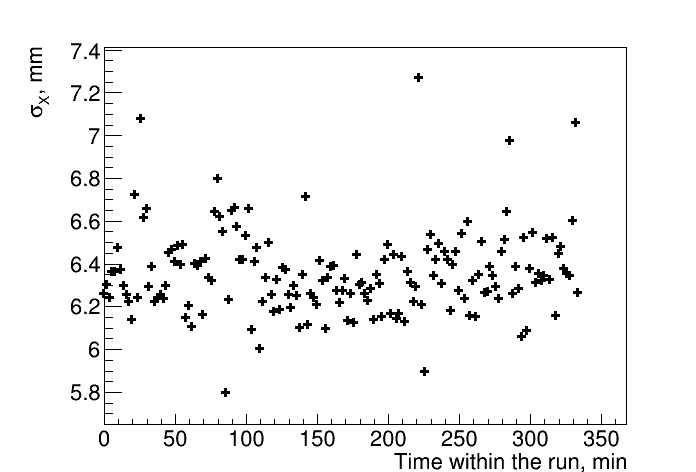}
    \end{minipage} \hfill
    \begin{minipage}{17.5pc}
      \centerline{
        \includegraphics[scale=0.3]{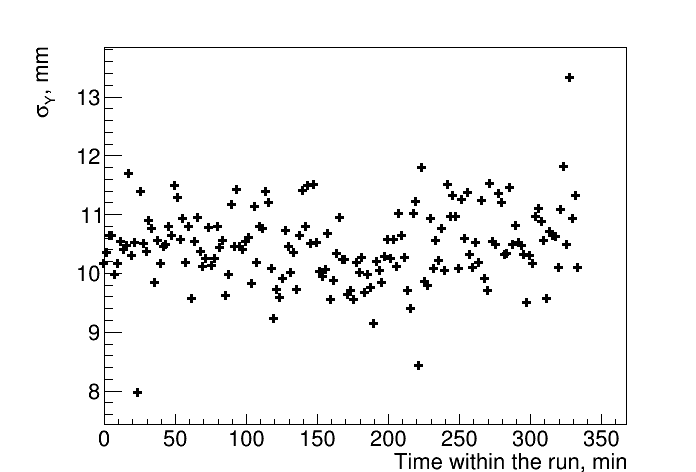}}
    \end{minipage}

 \ifdraft   
    \begin{minipage}{17.5pc}
      \centering
      \includegraphics[scale=0.3]{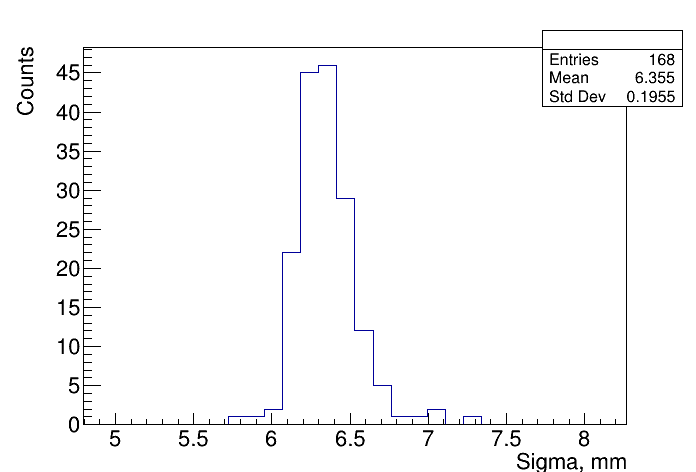}
    \end{minipage} \hfill
    \begin{minipage}{17.5pc}
      \centerline{
        \includegraphics[scale=0.3]{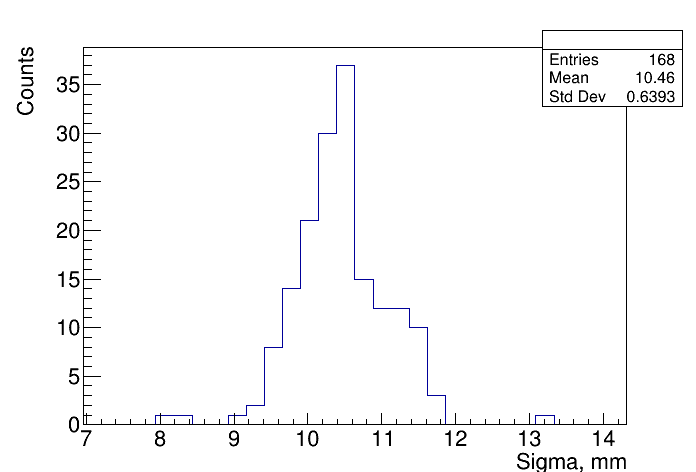}}
    \end{minipage}
\fi

    \caption{Typical variation of the 
    beam spread at the Timepix within a single run. 
    Left: X direction, right: Y direction.}
    \label{fig:timepix-single-run-sigma}
\end{figure}

Using the data for run 50444, 
the stability of the beam size during the a run with constant energy of the positrons
is shown in Fig. \ref{fig:timepix-single-run-sigma}. 

For both directions, the Gaussian sigma of the beam spot remained stable with negligible variation within the run. 
This is respected even in the case where a drift in the central position 
of the beam is observed (see Fig. \ref{fig:timepix-single-run}). 
The overall spread of the beam size was $\sim 200$~\textmu m in X direction and 
about 600~\textmu m in Y direction, 
where the difference was attributed to the lower statistics used for Y parameters determination.
   
\begin{figure}[h]
      \centering
      \includegraphics[scale=0.3]{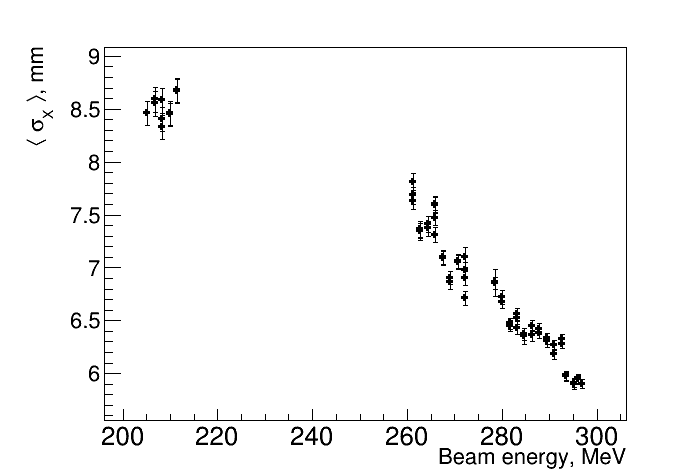}
    \caption{Correlation between the beam size in X direction at the Timepix plane
    and the beam energy.}
    \label{fig:timepix-sigma-vs-energy}
\end{figure}

The dependence of the beam spot size in X direction 
as a function of the beam 
energy during the period with stable 
Linac operating conditions is shown in fig. \ref{fig:timepix-sigma-vs-energy}. 
The beam size increases with the 
decrease of the beam energy due to the 
location of the Timepix downstream of the beam focal point. 
The beam size in Y direction is not dependent on the beam
energy and this is consistent with the 
configuration of the last focusing quadrupole
of the positron beam line. 

\section{Positron flux measurement}
\label{sec:Lumi}
The scan technique employed by PADME in Run III aims to discover X17 by observing deviations in the ratio $N_{2Cl}/N_{POT}$ as a function of the center of mass energy. Theoretical sensitivity estimates have shown that, for couplings of the order of $g_{ve}\sim 10^{-4}$, the enhancement in the ratio originated by X17 production  reaches the percent level.
Ensuring the accurate measurement of luminosity with a relative precision of $\sim$1\%, is therefore a crucial aspect. 

Since during most of the data taking in RUN III the Timepix was
operated in frame mode, the number of the recorded bunches 
within the acquisition window of 1 s varies within 2 \% 
(48 or 49 bunches recorded, depending on a random phase shift of 
the start of frame acquisition signal). 
This uncertainty prevents the usage of the Timepix for 
precise beam luminosity measurements. 
Still, the measurement of the multiplicity with Timepix 
is an additional valuable systematic check.

\begin{figure}[h]
\centering
    \includegraphics[width=8.1 cm]{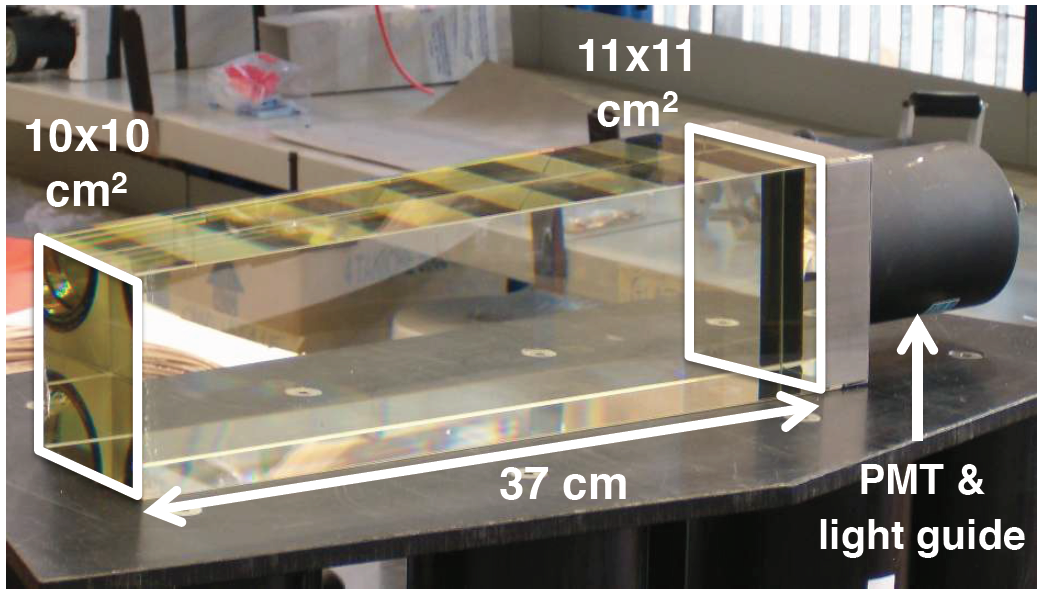}
    \caption{A module from the OPAL calorimeter without wrapping with approximate dimensions\cite{NA62:2017rwk}.}
    \label{fig:PbGL}
\end{figure}

Using the diamond active target as luminosity monitor a precision of $\sim$5\% was achieved in Run II, as described in \cite{PADME:2022tqr}.
The beam intensity used by PADME during Run III ($\sim$ 3 $\times10^3$ POT) with respect to Run II ($\sim$ 30$\times10^3$ POT) significantly reduced the precision of active target based measurements.
To improve the luminosity measurement precision, 
an SF57 lead glass block (PbGL) re-used from the OPAL experiment (see Fig. \ref{fig:PbGL}) has been installed in PADME during Run III. The block is positioned downstream of the ECal hole and the TimePix assembly, and is equipped with a single 2-inch fine-mesh Hamamatsu R2238 photomultiplier used for readout. The photomultiplier is connected to the trapezoidal prism block via a cylindrical SF57 light guide.
Details on the block characteristics can be found in \cite{NA62:2017rwk}. 

The PbGL block absorbs the beam particles and measures the total deposited energy. Following a meticulous calibration process, the accumulated charge obtained from the lead glass is appropriately scaled to the total deposited energy. Subsequently, knowing the beam energy $E_\mathrm{beam}$, and the fraction of energy collected by the PbGL block $F_{E}$ with respect to the one crossing the target, the total number of POT can be calculated using the formula:
\begin{equation}
N_{POT}^{bunch} = \frac{E_{\text{totLG}}}{E_\text{beam}F_{E}}
\label{eqn:NPOT}
\end{equation}
To calibrate the lead glass, a series of dedicated runs were collected at the beginning of October 2022.

The statistical error on the total number of POT for each point in the mass scan is entirely negligible. However, it is crucial to assess the stability of the detector and readout electronics at a level below 1\% for each energy value. This ensures that the outcomes are primarily influenced by statistical fluctuations rather than systematic uncertainties.
 
\subsection{The absolute luminosity calibration}

The luminosity measurement method is based on the determination of the amount of charge produced by the PbGL block for a single particle $Q_{1e}$ of known energy.  
In order to obtain the value of $E_\text{totLG}$ in \ref{eqn:NPOT} the accumulated charge in the PbGL needs to be converted in MeV. To compute the calibration constant, several special runs at known beam energy have been collected. 
Additionally, when increasing the number of particles, it is essential to verify the linearity of the detector. Special runs with varying high voltage (HV) on the PbGl block and different beam multiplicities have been carried out at a fixed beam energy value. These runs have demonstrated that the $N_{POT}$ measurement range can be extended to several thousand particles. All PbGL calibration runs were conducted at a nominal energy value of 402.5 MeV. Therefore, our focus will be on measuring the charge per single POT. Converting from charge $Q_{1e}$ to energy $Q_{1MeV}$ is straightforward; it involves dividing the $Q_{1e}$ by the deposited energy value.

Two independent methods have been used to measure the charge to energy conversion factor: 
\begin{itemize}
    \item fit to high multiplicity response normalised to BTF FitPix silicon pixel detector
     \item fit to the single particle response at 1000V extrapolated to 650 V using gain measurement
\end{itemize}
 
\subsubsection{Fit to high multiplicity}
The fit to the high multiplicity response is performed using a data sample collected in a $N_{POT}$ range from 200 to 3000. The block was operated at a high voltage value of 650V to prevent saturation of the readout electronics and nonlinear effects due to high current in the PMT divider.
These individual datasets were collected with the DHSTB002 magnet off, and the block was positioned in front of the dipole straight exit (see Fig. \ref{fig:selector}). The $N_{POT}$ estimate was provided by BTF using their silicon pixel sensor.
\begin{figure}[h]
\centering
    \includegraphics[width=7cm]{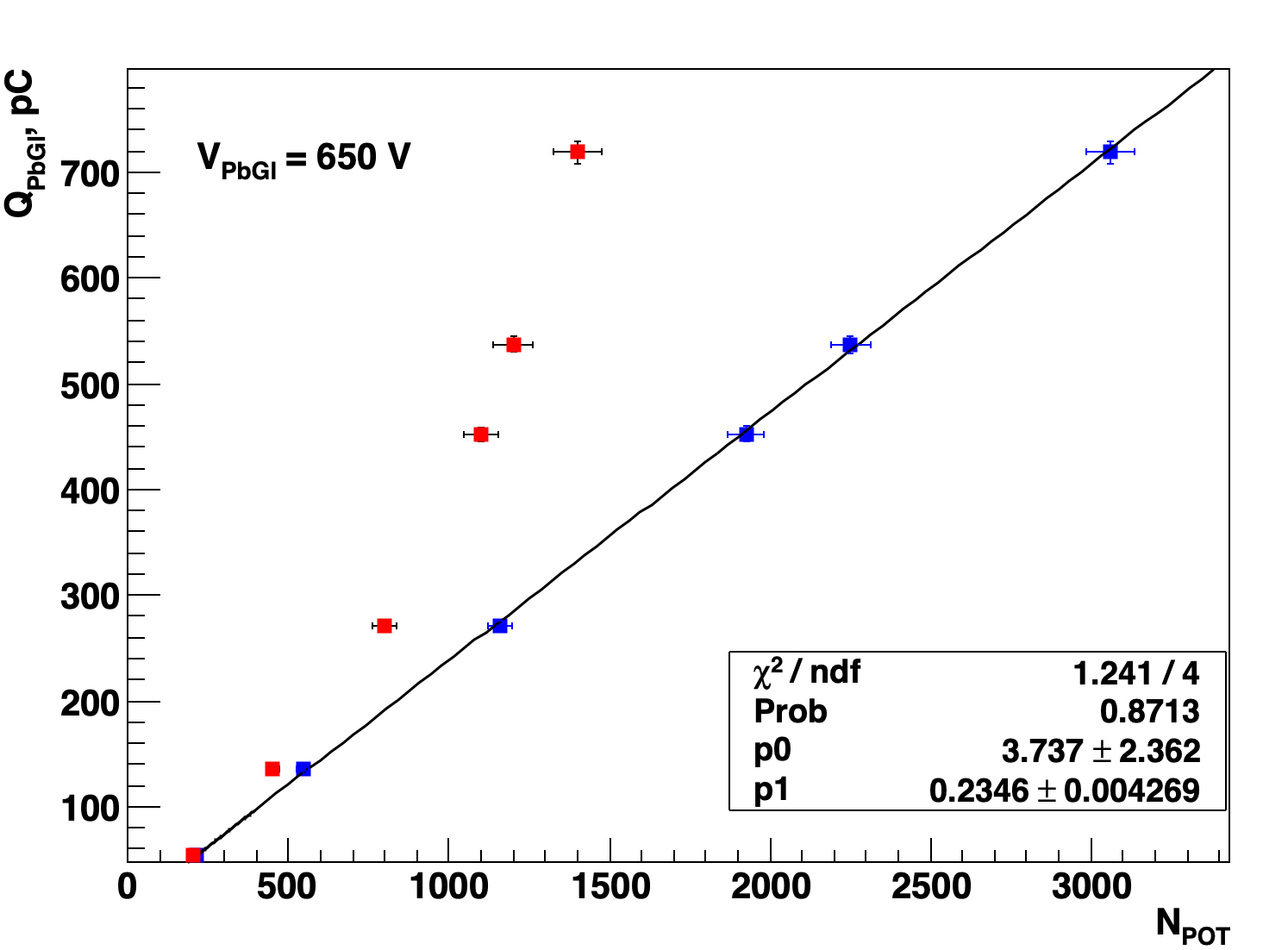}
    \caption{Calibration curve for the PbGL at a voltage setting of 650 V. Red point represent FitPix $N_{POT}$ values, blue point are obtained after applying Toy MC correction.}
    \label{fig:PbGL_CALIB_HIGH}
\end{figure}

The number of positron values provided by BTF, indicated by the red points in Fig. \ref{fig:PbGL_CALIB_HIGH}, deviated from linearity due to the high density of particles per mm$^2$ in the N$_{POT}>$500 region. This issue stemmed from the dead time of the silicon pixel detector, which was unable to count more than one particle per pixel. With a spot size as small as $0.5\times1.0$ mm$^2$, the likelihood of having more than one positron in the same pixel was far from remote in the high multiplicity region.

To correct for this effect a toy Monte Carlo has been developed able to simulate the spot size and provide the number of firing pixels as a function of the number of incoming particles. The simulation has been validated using few runs in which both silicon pixel based and an independent measurement of the bunch luminosity was provided by BTF luminometer.
\begin{figure}[h]
    \centering
    \begin{minipage}[t]{0.48\textwidth}
        \centering
        \includegraphics[width=\textwidth]{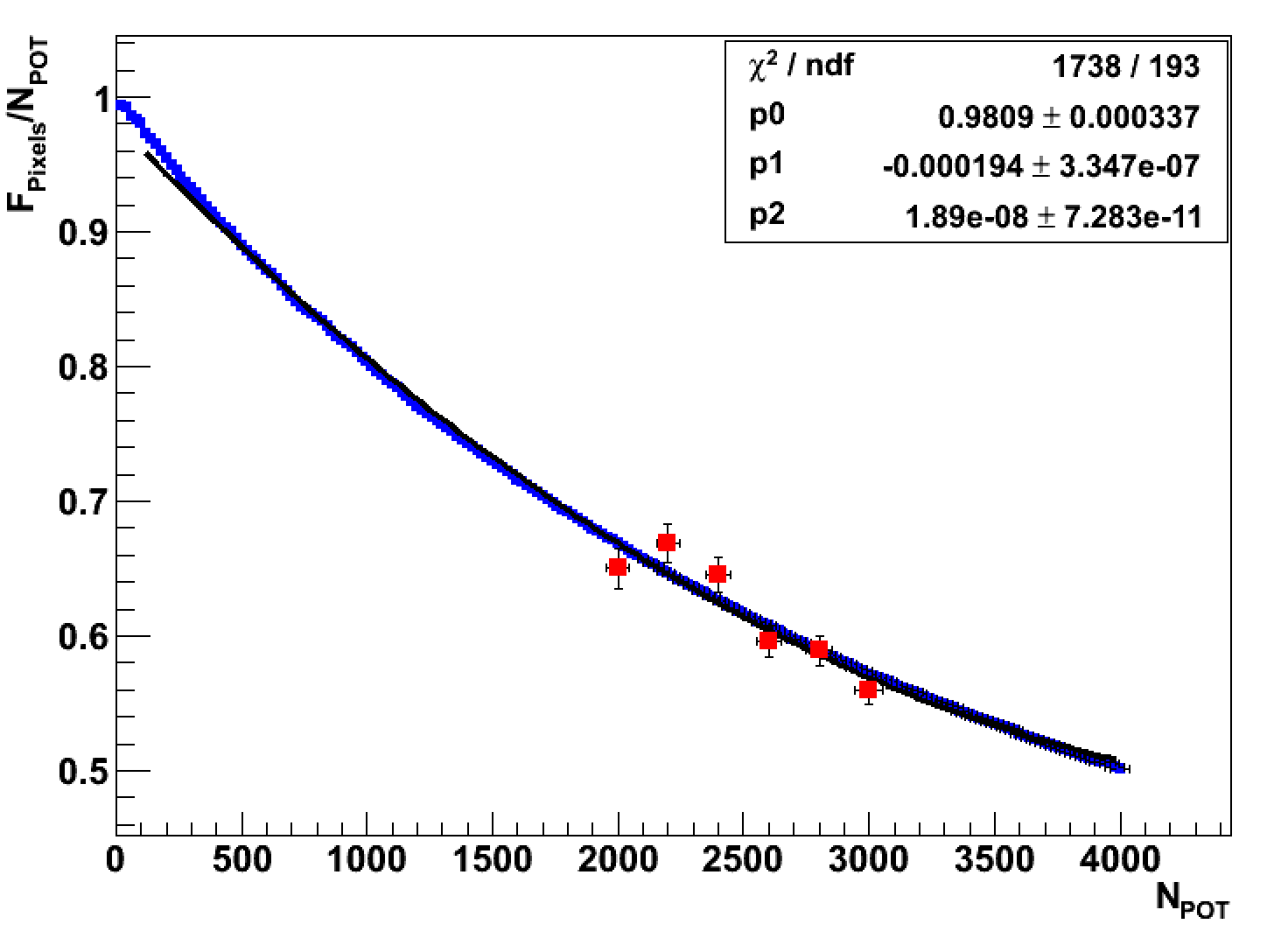}
        \caption{Toy MC fraction compared to BTF measurement.}
        \label{fig:FitPixCalib}
    \end{minipage}
    \hfill
    \begin{minipage}[t]{0.48\textwidth}
        \centering
        \includegraphics[width=\textwidth]{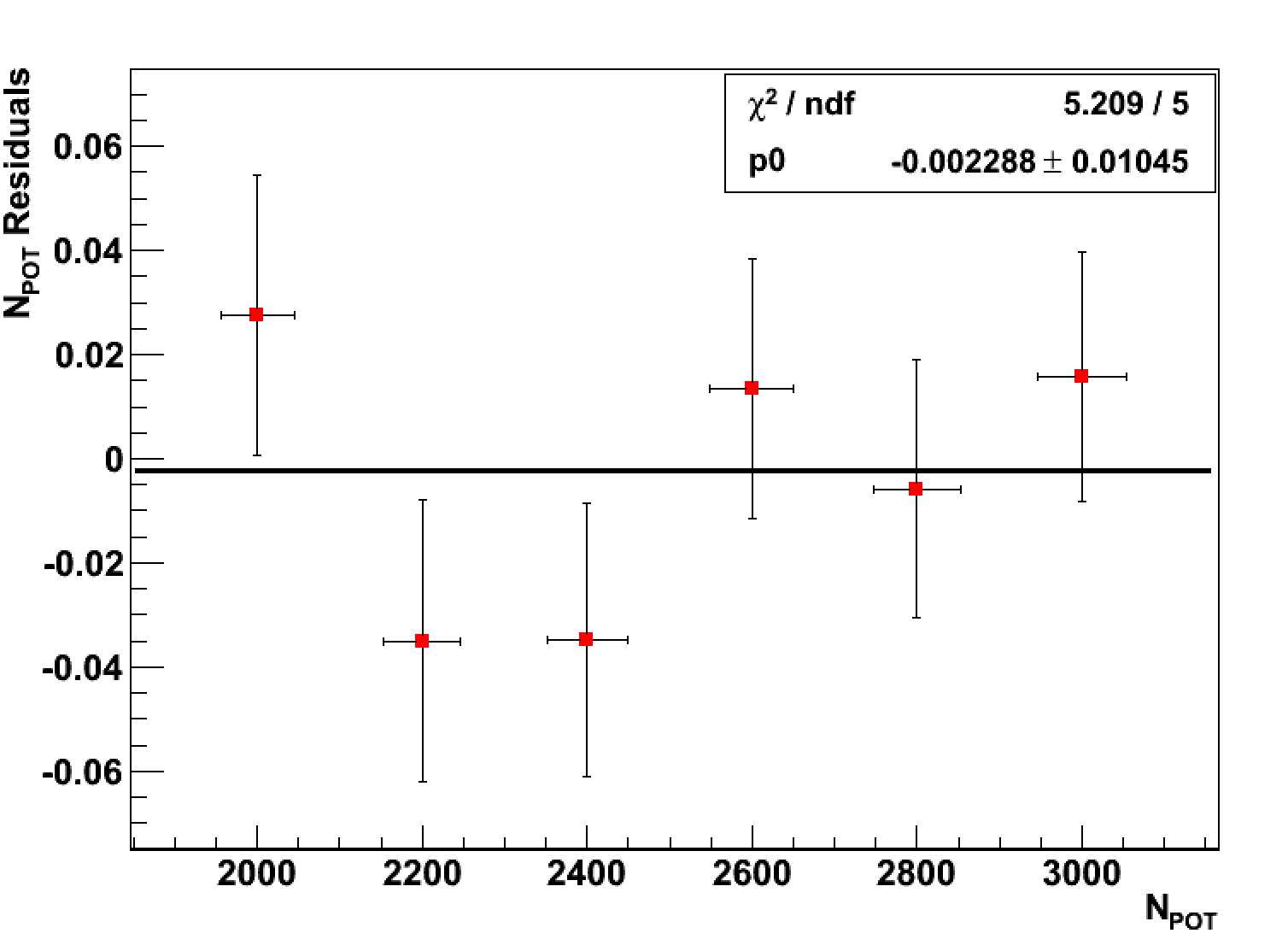}
        \caption{Residuals to the ToyMC prediction.}
        \label{fig:FitPixRes}
    \end{minipage}
\end{figure}

In Fig. \ref{fig:FitPixCalib}, the red points represent the ratio of the two numbers provided by BTF, based on the luminometer and the FitPix detector. These results are compared with the prediction of the Monte Carlo simulation, specifically the ratio between firing pixels $F_{pixels}$ and the number of simulated incident positrons. The Monte Carlo simulation utilizes the beam dimensions provided by the BTF FitPix detector as inputs, and the results are highly sensitive to it. Residuals, as shown in Fig. \ref{fig:FitPixRes}, demonstrate an extremely good agreement, proving that the non-linearity effect on FitPix was actually due to pile-up on the same pixel and can be efficiently corrected. Additionally, we can be confident that the FitPix $N_{POT}$ measurement scale is correct to a few percent level.  
The obtained values of the correction factors, adapted to the specific spot size used during the scan at 650 V, are applied to the red point in Fig. \ref{fig:PbGL_CALIB_HIGH} obtaining much better linear behavior shown by the blue points.
In fact the linear fit performed over all the $N_{POT}$ explored range exhibits a very good $\chi^2$. The obtained value for the single particle charge is $Q_{1e}=(0.235\pm0.0043)$ pC/particle. The fit shows very good linearity up to the 3000 POT/bunch which is the limit range explored by PADME during the Run III data taking, demonstrating that the PADME luminometer can be safely used at this HV values in a range from 30-3000 particle/bunch.

\subsubsection{Fit to single positron charge}
The extremely low charge per particle obtained with block high voltage setting 650 V, demonstrates that it's impossible to measure the single particle response at so low voltage. For this reason the high voltage setting was raised to 1000 V. According to Hamamatsu data sheet the gain increase should be $>$20-fold, enabling a measurable single positron charge value. Operating the block at 1000V during the data collection is also not feasible due to saturation and non-linearity effects caused by the excessively high collected charge in the regime of thousands of positrons per bunch.
The obtained $Q_{1e}(1000V)$ needs to be scaled down to 650V for comparison with the previously measured value based on FitPix calibration. Therefore, along with measuring the single particle response, it's necessary to extract the PMT gain curve to allow the scaling factor to be computed. 

\begin{figure}[h]
    \centering
    \begin{minipage}[t]{0.48\textwidth}
        \centering
        \includegraphics[width=\textwidth]{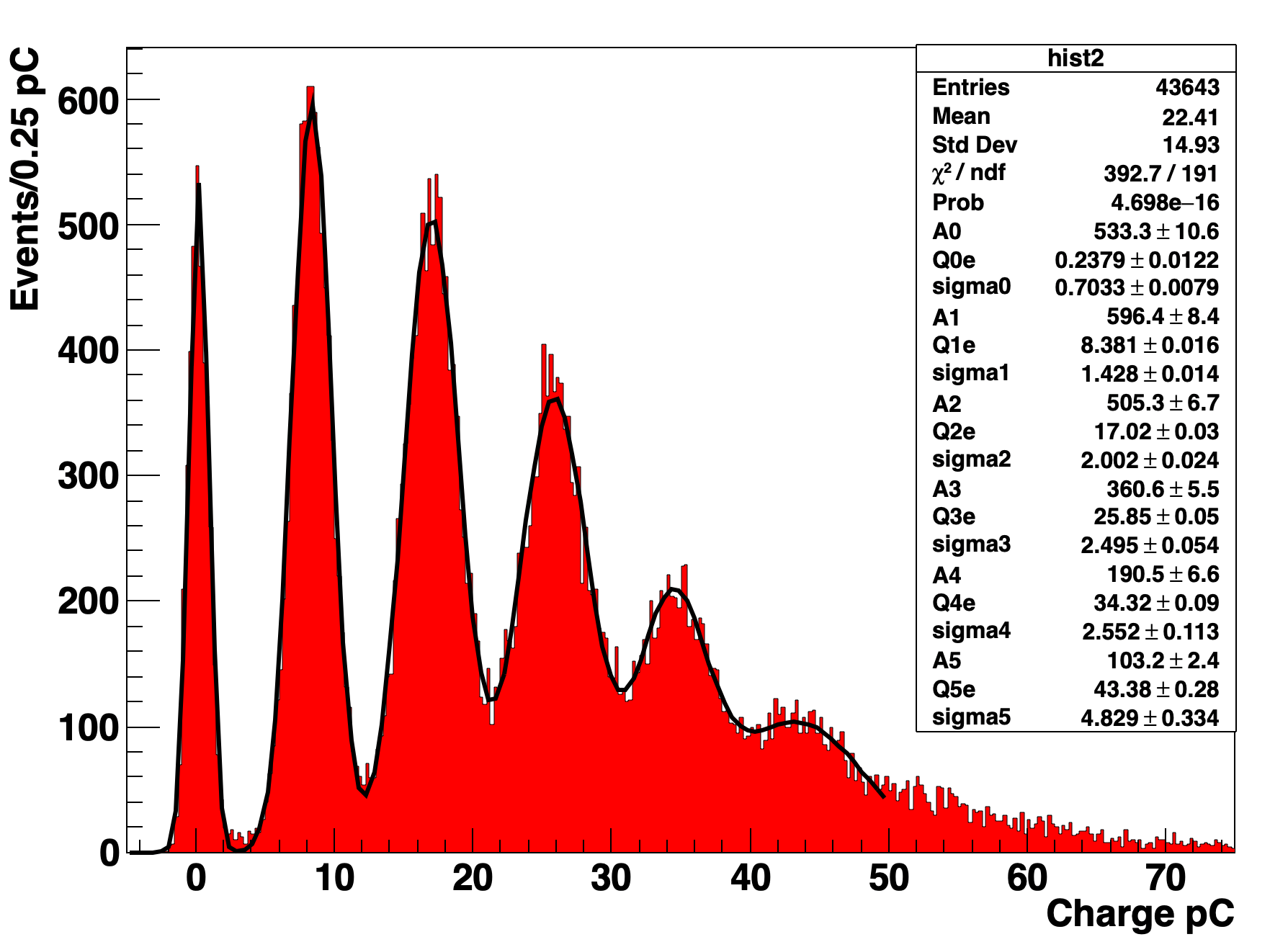}
        \caption{Single particle charge spectrum.}
        \label{fig:Q1eSpectrum}
    \end{minipage}
    \hfill
    \begin{minipage}[t]{0.48\textwidth}
        \centering
        \includegraphics[width=\textwidth]{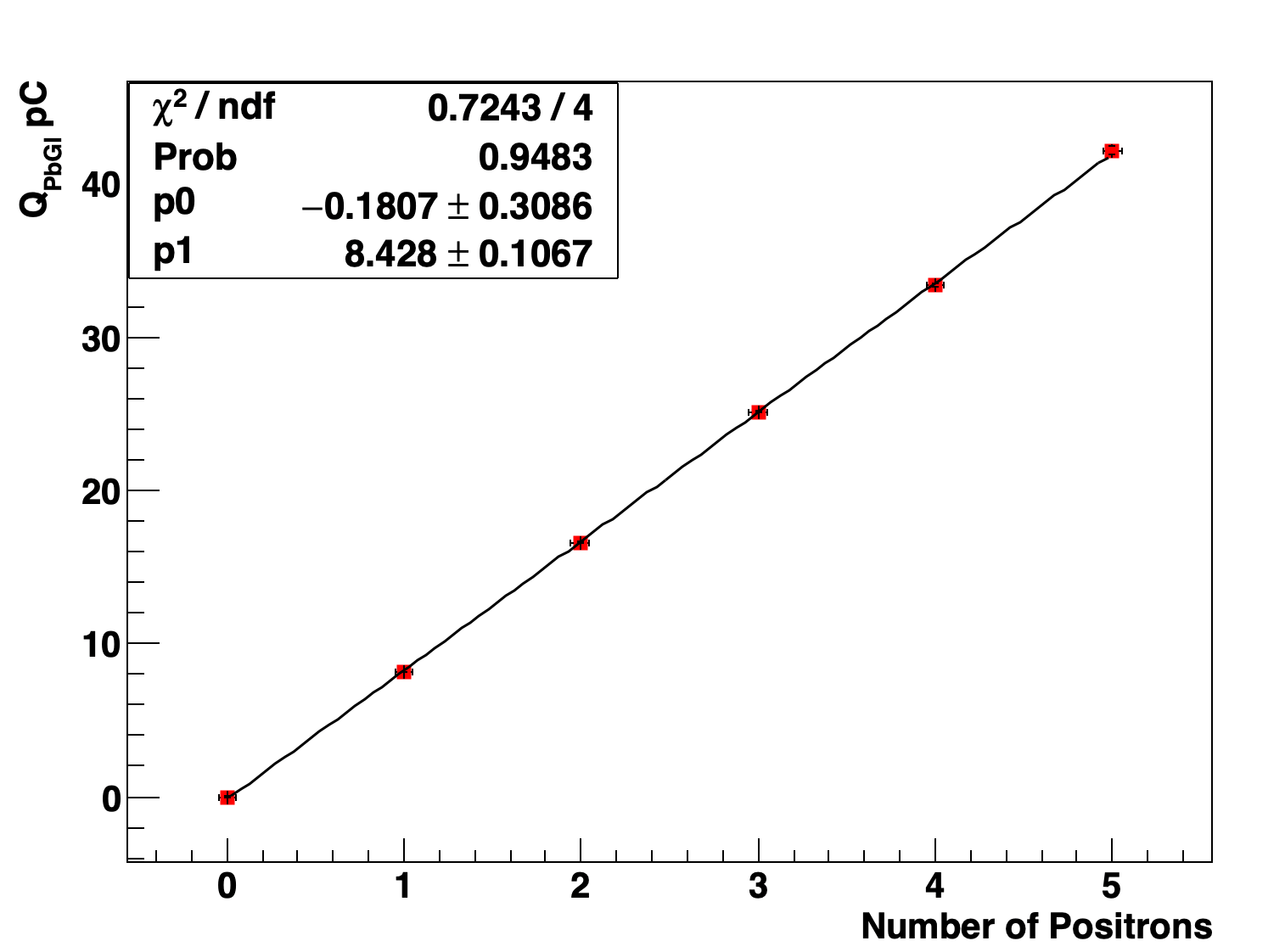}
        \caption{Fit to the single particle response.}
        \label{fig:Q1eFit}
    \end{minipage}
\end{figure}

The single positron response is obtained by fitting the Poissonian distribution of the spectrum acquired at 1000 V. Fig.~\ref{fig:Q1eSpectrum} displays the charge spectrum used which is fitted with the sum of 6 different independent Gaussian functions to extract the charge from 0 to 5 positrons. The fit for 0 electrons accounts for the pedestal and is subtracted from the all of the remaining multi electron peaks. The average values of the fit for each gaussian are shown in Fig.\ref{fig:Q1eFit} together with a linear fit. The single particle charge value obtained at 1000 V is $Q_{1e}^{1000}=(8.384\pm0.035)$ pC with a relative error $<$0.5\%. 

Comparing with the direct fit to single particles in Fig. \ref{fig:Q1eSpectrum}, after subtracting the pedestal and summing up the errors, we obtain $Q_{1e}=(8.143\pm0.028)$ pC. The discrepancy of central values is approximately 2.8\%. Since the errors are comparable, we will use the weighted average as the determination of $Q_{1e}^{1000}$ and assign a 1.4\% systematic error. Our best determination of $Q_{1e}^{1000}$ will then be:
$Q_{1e}^{1000}=(8.24 \pm 0.021_{Stat} \pm 0.12_{Syst})$ pC.

\subsubsection{The PbGL gain measurement}

Moving to the final step, the scaling factor for comparing $Q_{1e}^{1000}$ to that at 650V $Q_{1e}^{650}$needs to be determined. For this purpose, PADME has collected a series of 4 different runs in single particle mode with varying the PbGL high voltage setting values, namely 950V, 1000V, 1050V, and 1100V. For each of these runs, we fit the single particle response spectrum, similar to the one in Fig. \ref{fig:Q1eSpectrum},
to generate the charge vs. high voltage curve. To scale the charge to a gain curve we used the value of the light yield of the block measured by NA62 in 2014 $LY_{NA62}=0.331\pm 0.006$ photoelectrons/MeV\cite{NA62:2017rwk}. The value of the gain is obtained for each HV value and $N_{pos}$ using the formula:
\begin{equation}
G(HV,N_{pos})=\frac{Q_{1e}(HV,N_{pos})}{e*LY_{NA62}*E_{Beam}*N_{pos}*Frac_{Edep}}  
\label{eqn:GainScale}
\end{equation}
where $Frac_{Edep}$ represents the fraction of energy deposited in the PbGL by a positron of energy $E=E_{Beam}$. This fraction is determined through Monte Carlo simulations and is independent of energy for a central impact position.
\begin{figure}[h]
        \centering
        \includegraphics[width=0.8\textwidth]{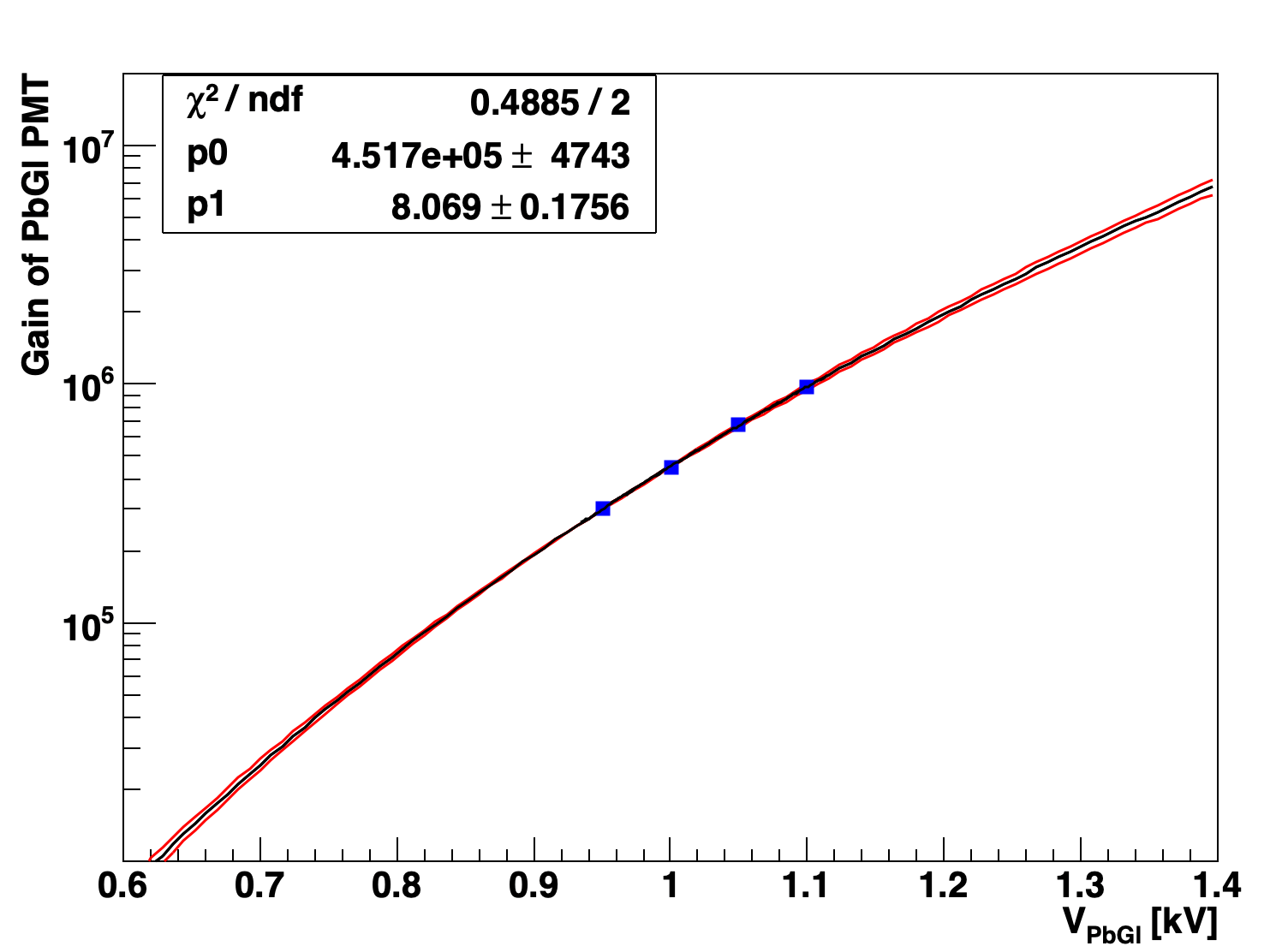}
        \caption{Gain curve with error range in red.}
        \label{fig:GainScale}
\end{figure}

In Fig.\ref{fig:GainScale}, the gain curve obtained by scaling $Q_{1e}$ according to Eq. \ref{eqn:GainScale} for each voltage setting is shown. The curve is approximated with the function provided by Hamamatsu:
\begin{equation}
    G(HV)=p_{0} \times V^{p_1}=G_{1000} \times V^{kn}
    \label{eqn:GainFunc}
\end{equation}
with V in kV.
The value of $p_1=kn$, predicted by Hamamatsu to fall within the range of 8.4-9.6, closely matches the value we obtained. Additionally, we attempted to fit the 2 and 3 positron peaks, resulting in compatible outcomes. 
The red curves represent the limits for $\pm1 \sigma$ varying the parameter including correlations. From the distribution of obtained values of the gain at 650 V we obtained $G(650V)=(14001\pm1121)$ with an error of ±8.0\%. The ratio of $G(1000V)$ to $G(650V)$ is then determined to be: $\frac{G(1000)}{G(650)}=32.26\pm 2.6$.

We emphasize that the normalization to the gain provided by Eq. \ref{eqn:GainScale} does not influence the procedure, and the scaling factor in Eq. \ref{eqn:GainFunc} could be directly obtained by fitting the charge $Q_1e(V)$. In the future, reducing the error in the extrapolation could be achieved by measuring the gain for voltage settings closer to 650V, and by increasing the granularity of the voltage scan to 25V steps.
 
\subsubsection{Comparing results}
Now that the gain at 650V has been measured, we can compare the charge values obtained using the two different methods, which are briefly summarized.

In the first method, known as the multi-particle fit, we fit the curve of charge versus $N_{POT}$ at 650V, where $N_{POT}$ is provided by the BTF FitPix detector. We developed a re-scaling factor for the FitPix $N_{POT}$ measured value based on Monte Carlo simulations, which has been shown to improve linearity and is in agreement with the BTF calorimetric-based calibration to within approximately 1\%.    
In the second method, known as the single-particle fit, we fit the single-particle response at 1000V for up to 5 positrons, and we use the slope parameter to determine $Q_{1e}$(1000). Subsequently, we measure the PMT gain at 650V by extrapolating the gain curve, and obtain $Q_{1e}$(650) by scaling $Q_{1e}$(1000) using the gain ratio. We get: $Q_{1e}$(650)=$Q_{1e}$(1000)/GRatio.
The two methods are completely independent procedures and utilize separate data samples, thus any differences between them account for systematic errors inherent to each method.
Tab.\ref{tab:lumiRecap} summarizes the input used in the calculation and the values obtained for $Q_{1e}^{650}$.

\begin{table}[h]
\centering
\caption{Comparison of $Q_{1e}^{650}$ results and errors}
\label{tab:results}
\begin{tabular}{lccc||lccc}
\hline
 & \multicolumn{3}{c}{single-particle} & \multicolumn{4}{c}{multi-particle} \\
& Value & Error & Error \% &      & Value & Error& Error \% \\
\hline
$Q_{1e}^{1000}$ & 8.24 & 0.122 & 1.46 & $Q_{1e}^{650}$ & 0.235 & 0.0042& 1.8 \\
Gain ratio & 32.26 & 2.6 & 8. &FitPix scale & &0.0024&1.0 \\
\hline
$Q_{1e}^{650}$ & 0.255 & 0.021 &8.2 &$Q_{1e}^{650}$ & 0.235 & 0.0046&2.0 \\
\hline
\label{tab:lumiRecap}
\end{tabular}
\end{table}

The obtained values are compatible, differing by just 8.3\%, which corresponds to $\sim$1.$\sigma$. However, the single particle fit exhibits a fourfold higher uncertainty, dominated by the gain extrapolation procedure. For this reason we will use the multi particle fit as determination of the calibration factor $Q_{1e}^{650}$ in the PADME luminosity measurement.

\subsection{Corrections for beam conditions variation}

During data collection, the PADME PbGL beam monitor was relocated from its original position at the straight exit of DHSTB002, where it underwent calibration, to the end of the PADME experimental setup as depicted in Fig. \ref{fig:selector}. This change in positioning raises the possibility of beam conditions affecting either the PbGL response or the energy reaching it. As a result, several aspects require careful consideration:
\begin{itemize}
    \item Discrepancies in the collected energy resulting from different beam energy values.
    \item Discrepancies in the energy leakage due to varying beam directions.
    \item Discrepancies in the energy leakage due to differences in beam spot dimensions.
    \item Variations in the collected energy caused by the beam's energy loss in the PADME detector.
\end{itemize}

These effects are expected to be small, they are nevertheless simulated in the PADME Monte Carlo, and corrections to the positron flux measurement are subsequently obtained for each energy point explored. These second-order corrections will not be addressed in this paper because they depend on specific beam conditions. 

\subsection{Check on the off resonance data sample}

To demonstrate that the $N_{POT}$ measurement achieves the desired relative sensitivity (approximately 1\%), we utilized a dataset collected with a beam energy of 402.5 MeV. PADME conducted a series of 5 distinct runs, and we anticipate measuring an identical value for the ratio $N_{2Cl}/N_{POT}$ in each run, provided that the number of positrons on target is accurately evaluated. The accumulated $N_{POT}$ ranges from $1.5\times10^9$ to $5.6\times10^9$, providing a number 2 clusters in the range $4.5\times10^3$ to $16\times10^3$ allowing to test sensitivities down to the 1\% level.

\begin{figure}[h]
\centering
    \includegraphics[width=8.1 cm]{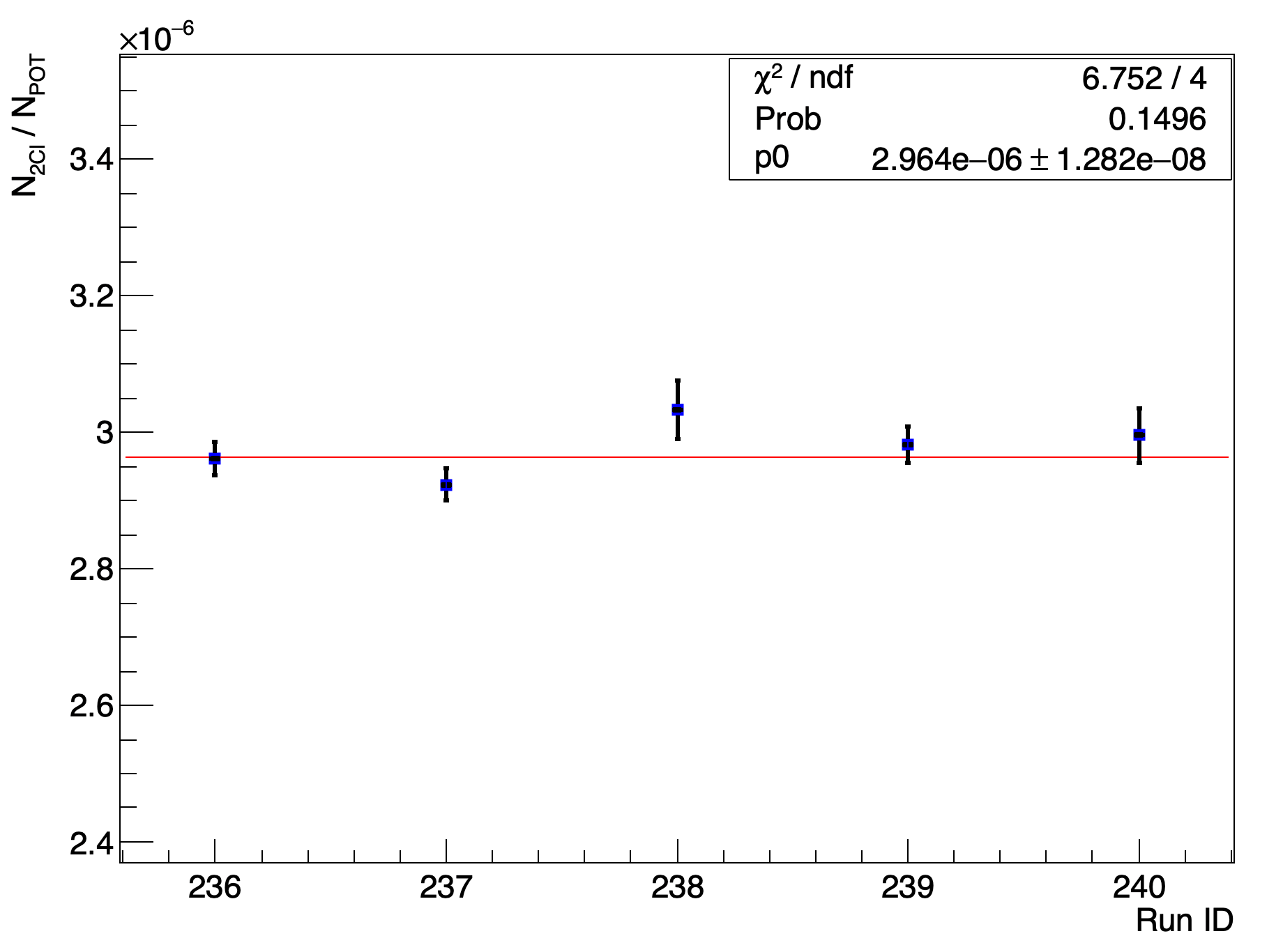}
    \caption{values of $N_{2Cl}/N_{POT}$ vs Run number for the 5 different runs.}
    \label{fig:OvRes}
\end{figure}
Fig. \ref{fig:OvRes} shows the stability of the ratio over the 5 different runs, indicated on the X axis. 
The measured values agree very well with the constant fit hypothesis, and the good $\chi^2$ demonstrates the absence of per cent scale systematic errors due to inconsistent $N_{POT}$ measurements. We emphasize that the data set has been collected on a 2 day time scale and in very stable beam conditions.

\section{Conclusions}
\label{sec:Conc}

This paper outlines the studies conducted by the PADME collaboration to determine the main characteristics of the beam delivered by the BTF during Run III, which took place from October to December 2022. The document demonstrates the capability of the PADME apparatus to monitor the absolute beam momentum to a precision of 1-2 MeV, the beam position to the level of few mm, and the luminosity to a precision of 2\% level absolute and to $<$1\% relative. Finally the beam energy spread has been estimated to be to lower than 0.25\%. 

The error in the absolute momentum scale corresponds to an uncertainty of approximately 40 KeV on the X17 mass scale, while the 2\% error in the absolute luminosity measurement results in a 1\% uncertainty in its coupling strength to electrons. Furthermore, the energy spread value of less than 0.25\% renders the beam contribution to production rates negligible compared to the effects of electron motion \cite{Arias-Aragon:2024qji}. 
Finally, a relative luminosity measurement stability of better than 1\%  over a 2-day time scale has been demonstrated by utilizing off-resonance data.
This aligns with the expectations of the PADME experiment, positioning it to offer precise exclusions within the X17 parameter space.

\section*{Acknowledgments}

The PADME collaboration acknowledges significant support from the Istituto Nazionale di Fisica Nucleare, and in particular the Accelerator Division and the LINAC team of INFN Laboratori Nazionali di Frascati, for providing an excellent quality beam and full support during the data collection period. The authors are also grateful to E. Nardi, L. Darmè, and G. Grilli di Cortona for the enlightening discussions regarding the X17 search strategy.
Sofia University team acknowledges that this study is partially financed by the European Union-NextGenerationEU,
through the National Recovery and Resilience Plan of the Republic of Bulgaria,
project SUMMIT BG-RRP-2.004-0008-C01 and 
TA-LNF as part of STRONG-2020 EU Grant Agreement 824093.

%% file: main.bbl
\providecommand{\href}[2]{#2}\begingroup\raggedright\begin{thebibliography}{10}

\bibitem{Holdom:1986eq}
B.~Holdom, \emph{{Searching for $\epsilon$ Charges and a New U(1)}}, \href{https://doi.org/10.1016/0370-2693(86)90470-3}{\emph{Phys. Lett. B} {\bfseries 178} (1986) 65}.

\bibitem{PADME:2022fuc}
{\scshape PADME} collaboration, \emph{{Commissioning of the PADME experiment with a positron beam}}, \href{https://doi.org/10.1088/1748-0221/17/08/P08032}{\emph{JINST} {\bfseries 17} (2022) P08032} [\href{https://arxiv.org/abs/2205.03430}{{\ttfamily 2205.03430}}].

\bibitem{Antel:2023hkf}
C.~Antel et~al., \emph{{Feebly Interacting Particles: FIPs 2022 workshop report}},  in \emph{{Workshop on Feebly-Interacting Particles}}, 5, 2023 [\href{https://arxiv.org/abs/2305.01715}{{\ttfamily 2305.01715}}].

\bibitem{Krasznahorkay:2015iga}
A.J.~Krasznahorkay et~al., \emph{{Observation of Anomalous Internal Pair Creation in Be8 : A Possible Indication of a Light, Neutral Boson}}, \href{https://doi.org/10.1103/PhysRevLett.116.042501}{\emph{Phys. Rev. Lett.} {\bfseries 116} (2016) 042501} [\href{https://arxiv.org/abs/1504.01527}{{\ttfamily 1504.01527}}].

\bibitem{Krasznahorkay:2021joi}
A.J.~Krasznahorkay, M.~Csatl\'os, L.~Csige, J.~Guly\'as, A.~Krasznahorkay, B.M.~Nyak\'o et~al., \emph{{New anomaly observed in He4 supports the existence of the hypothetical X17 particle}}, \href{https://doi.org/10.1103/PhysRevC.104.044003}{\emph{Phys. Rev. C} {\bfseries 104} (2021) 044003} [\href{https://arxiv.org/abs/2104.10075}{{\ttfamily 2104.10075}}].

\bibitem{Krasznahorkay:2022pxs}
A.J.~Krasznahorkay et~al., \emph{{New anomaly observed in C12 supports the existence and the vector character of the hypothetical X17 boson}}, \href{https://doi.org/10.1103/PhysRevC.106.L061601}{\emph{Phys. Rev. C} {\bfseries 106} (2022) L061601} [\href{https://arxiv.org/abs/2209.10795}{{\ttfamily 2209.10795}}].

\bibitem{Zhang:2017zap}
X.~Zhang and G.A.~Miller, \emph{{Can nuclear physics explain the anomaly observed in the internal pair production in the Beryllium-8 nucleus?}}, \href{https://doi.org/10.1016/j.physletb.2017.08.013}{\emph{Phys. Lett. B} {\bfseries 773} (2017) 159} [\href{https://arxiv.org/abs/1703.04588}{{\ttfamily 1703.04588}}].

\bibitem{Alves:2023ree}
D.S.M.~Alves et~al., \emph{{Shedding light on X17: community report}}, \href{https://doi.org/10.1140/epjc/s10052-023-11271-x}{\emph{Eur. Phys. J. C} {\bfseries 83} (2023) 230}.

\bibitem{Nardi:2018cxi}
E.~Nardi, C.D.R.~Carvajal, A.~Ghoshal, D.~Meloni and M.~Raggi, \emph{{Resonant production of dark photons in positron beam dump experiments}}, \href{https://doi.org/10.1103/PhysRevD.97.095004}{\emph{Phys. Rev. D} {\bfseries 97} (2018) 095004} [\href{https://arxiv.org/abs/1802.04756}{{\ttfamily 1802.04756}}].

\bibitem{Darme:2022zfw}
L.~Darm\'e, M.~Mancini, E.~Nardi and M.~Raggi, \emph{{Resonant search for the X17 boson at PADME}}, \href{https://doi.org/10.1103/PhysRevD.106.115036}{\emph{Phys. Rev. D} {\bfseries 106} (2022) 115036} [\href{https://arxiv.org/abs/2209.09261}{{\ttfamily 2209.09261}}].

\bibitem{Ghigo:2003gy}
A.~Ghigo, G.~Mazzitelli, F.~Sannibale, P.~Valente and G.~Vignola, \emph{{Commissioning of the DAFNE beam test facility}}, \href{https://doi.org/10.1016/j.nima.2003.07.017}{\emph{Nucl. Instrum. Meth. A} {\bfseries 515} (2003) 524}.

\bibitem{DAFNE_site}
``{DA$\Phi$NE web site}.''.

\bibitem{Buonomo:2015pba}
B.~Buonomo, L.G.~Foggetta and G.~Piermarini, \emph{{New Gun Implementation and Performance of the DA\ensuremath{\Phi}NE LINAC}},  in \emph{{6th International Particle Accelerator Conference}}, p.~TUPWA056, 2015, \href{https://doi.org/10.18429/JACoW-IPAC2015-TUPWA056}{DOI}.

\bibitem{PADME:2022ysa}
{\scshape PADME} collaboration, \emph{{The PADME beam line Monte Carlo simulation}}, \href{https://doi.org/10.1007/JHEP09(2022)233}{\emph{JHEP} {\bfseries 09} (2022) 233} [\href{https://arxiv.org/abs/2204.05616}{{\ttfamily 2204.05616}}].

\bibitem{Foggetta:2021gdg}
L.~Foggetta et~al., \emph{{The Extended Operative Range of the LNF LINAC and BTF Facilities}},  in \emph{{12th International Particle Accelerator Conference~}}, 8, 2021, \href{https://doi.org/10.18429/JACoW-IPAC2021-THPAB113}{DOI}.

\bibitem{Arias-Aragon:2024qji}
F.~Arias-Arag\'on, L.~Darm\'e, G.G.~di~Cortona and E.~Nardi, \emph{{Production of dark sector particles via resonant positron annihilation on atomic electrons}},  \href{https://arxiv.org/abs/2403.15387}{{\ttfamily 2403.15387}}.

\bibitem{timepix:padme-beam}
S.~Bertelli et~al., \emph{Beam diagnostics with silicon pixel detector array at padme experiment}, \href{https://doi.org/10.1088/1748-0221/19/01/C01016}{\emph{Journal of Instrumentation} {\bfseries 19} (2024) C01016}.

\bibitem{PADME:2022tqr}
{\scshape PADME} collaboration, \emph{{Cross-section measurement of two-photon in-flight annihilation of positrons at $\sqrt{s}=20$\,MeV with the PADME detector}}, \href{https://doi.org/10.1103/PhysRevD.107.012008}{\emph{Phys. Rev. D} {\bfseries 107} (2023) 012008} [\href{https://arxiv.org/abs/2210.14603}{{\ttfamily 2210.14603}}].

\bibitem{NA62:2017rwk}
{\scshape NA62} collaboration, \emph{{The Beam and detector of the NA62 experiment at CERN}}, \href{https://doi.org/10.1088/1748-0221/12/05/P05025}{\emph{JINST} {\bfseries 12} (2017) P05025} [\href{https://arxiv.org/abs/1703.08501}{{\ttfamily 1703.08501}}].

\end{thebibliography}\endgroup
